\documentclass[twocolumn]{aastex702}

\newcommand{\Msun}{\,{\rm M_\odot}}
\newcommand{\fedd}{\,{f_{\rm Edd}}}
\newcommand{\Mblack}{M_\bullet}
\newcommand{\Mstar}{M_\star}

\newcommand{\rowlab}[1]{\cellcolor{tabnavy!4}\makebox[0.36\textwidth][l]{\hspace*{0.6em}#1}}

\newcommand{\secrow}[1]{\hspace*{-\tabcolsep}{\setlength{\fboxsep}{0pt}%
  \colorbox{tabnavy}{\rule[-0.9ex]{0pt}{3.4ex}\makebox[\dimexpr\textwidth-2\tabcolsep\relax][l]{%
  \hspace*{0.8em}\textcolor{white}{\textbf{#1}}}}}}

\usepackage{colortbl}
\definecolor{tabnavy}{HTML}{1F2A55}
\definecolor{waltzcol}{HTML}{3D3B8E}
\definecolor{tangocol}{HTML}{F28E2B}

\shorttitle{The Dance of the Little Red Dots}
\shortauthors{Pacucci \& Rinaldi}

\begin{document}

\title{The Dance of the Little Red Dots: Spectroscopy of Two Blue Companions at {\boldmath $z = 3.9$} and {\boldmath $5.5$}\\ Reveals Young, Metal-Poor Starbursts Merging with Their LRDs within a Few Hundred Myr}

\correspondingauthor{Fabio Pacucci} 

\author[orcid=0000-0001-9879-7780]{Fabio Pacucci}
\affiliation{Center for Astrophysics $\vert$ Harvard \& Smithsonian, 60 Garden St, Cambridge, MA 02138, USA}
\email[show]{fabio.pacucci@cfa.harvard.edu}

\author[orcid=0000-0002-5104-8245]{Pierluigi Rinaldi}
\affiliation{Department of Astronomy, University of Texas, Austin, TX 78712, USA}
\affiliation{Cosmic Frontier Center, The University of Texas at Austin, Austin, TX 78712, USA}
\email{prinaldi@utexas.edu}

\begin{abstract}
The Little Red Dots (LRDs) are compact, red sources, abundant at $4 \lesssim z \lesssim 9$ and likely powered by accreting black holes. Photometric studies have found blue companions around them, but their nature remains spectroscopically untested. This Letter presents the only two LRD--companion pairs with spectroscopy of both members: ``Waltz'' at $z = 3.9$ and ``Tango'' at $z = 5.5$, with projected separations of $3.6$ and $2.5$ kpc. Both companions lie at the redshift of their LRD, within $132$ and $57 \, \rm km \, s^{-1}$, and, though $524$ Myr apart, are physically similar: starbursts $\lesssim 100$ Myr old, metal-poor ($0.2 \, \rm Z_\odot$), dust-poor, dynamically cold, as bright as the LRDs in the rest-frame UV, and linked to them by diffuse emission. The LRD in Waltz hosts a black hole of virial mass $\approx 3 \times 10^{6} \Msun$ accreting at half the Eddington rate, a few times the birth mass of a direct collapse black hole, and embedded in gas as metal-poor as its companion. Orbital modeling finds Tango bound in $71-90\%$ of realizations across the halo-mass prior, and Waltz likely bound only if its host halo exceeds $5 \times 10^{10} \Msun$; bound pairs merge within median times of $129$ and $457$ Myr. We conclude that we are likely observing what heavy seeding requires: a young black hole beside the metal-poor starburst that the synchronized-pair scenario predicts, in the last steps of a cosmic dance that may end in a merger, and in a galaxy seeded by its black hole.
\end{abstract}


\section{Introduction}
\label{sec:introduction}

The Little Red Dots (LRDs) are compact, red sources discovered by the \textit{James Webb Space Telescope} (JWST) in its first deep-field campaigns \citep{Kocevski_2023, Harikane_2023, Matthee_2023, Maiolino_2023_new}. Photometric censuses now count hundreds of LRDs over the redshift range $4 \lesssim z \lesssim 9$ \citep{Kokorev_2024_census, Kocevski_2024, Akins_2025, rinaldi_way_2026}, suggesting an average surface density of $\approx1 \, \rm arcmin^{-2}$ \citep{rinaldi_way_2026, Perez-Gonzalez_2026}. 

Two photometric properties generally define the class: (i) compactness at optical wavelengths (\citealt{Hviding_2026, rinaldi_way_2026}), with typical effective radii of $\lesssim 100$ pc \citep{Baggen_2023}, and (ii) a V-shaped spectral energy distribution (SED), characterized by a blue UV and a red optical continuum with, on average, a turnover near the Balmer limit \citep{Setton_2024, Kocevski_2024}.

Spectroscopy has complicated the picture. Broad Balmer lines, with widths of $1000-3000 \, \rm km \, s^{-1}$, are detected in a large fraction of the LRDs observed with NIRSpec \citep{Greene_2023, Killi_2024, Taylor_2024}, and the same spectra frequently show Balmer absorption, which requires dense gas along the line of sight \citep{Deugenio_2025, juodzbalis_incomplete_2026}. If those lines are virialized, the implied black hole masses lie $1-2$ orders of magnitude above the local expectation for the stellar mass of the host \citep{Pacucci_2023_JWST, Jones_2025, Gupta_2026}. However, the X-ray emission such black holes should produce is rarely detected \citep{Maiolino_2024_Xray, Pacucci_Narayan_2024, Madau_2024, Lambrides_2024}. The virial black hole mass estimate is also under scrutiny, since electron scattering in a dense envelope can broaden a narrow line and inflate the inferred mass, potentially leading to an overestimate \citep{Rusakov_2025, Modelevsky_2026}.

The most basic question is still unanswered: {\it what powers an LRD?} Explanations divide on whether an accreting black hole is present at all. For instance, using radiation-hydrodynamic simulations, \cite{Pacucci_2026_DCBH} showed that the population's observational signatures follow if LRDs are powered by accreting direct collapse black holes (DCBHs) inside their natal halos. A second interpretation attributes the reddening to gas immediately around the black hole, as in the ``black hole star'' picture \citep{Naidu_2025a}, or in the quasi-star picture, a stage on the direct collapse pathway in which the black hole accretes from a massive envelope \citep{Begelman_2026, Santarelli_2026, Gentile_2026}. If no black hole is assumed, the LRD is a different kind of system: for example, a globular cluster in formation \citep{Chisholm_2026} or a single supermassive star \citep{Nandal_2026}. Of these models, direct collapse places demands on an LRD's immediate surroundings.

Direct collapse \citep{Loeb_Rasio_1994, Oh_Haiman_2002, Bromm_Loeb_2003, Begelman_2006, Lodato_Natarajan_2006} produces seeds with initial masses of $10^{4}-10^{6} \Msun$ \citep{Ferrara_2014}. The collapse proceeds only where: (i) the gas is pristine, (ii) the halo has reached the atomic-cooling threshold, and (iii) an external flux of Lyman-Werner (LW) photons ($11.2-13.6$ eV; \citealt{Draine_1996}) dissociates H$_2$ and suppresses fragmentation \citep{Omukai_2001, Latif_2014_UV, Inayoshi_review_2019}. Since the flux falls rapidly with increasing distance, the irradiating source must be a near neighbor. 

In the ``synchronized pair'' scenario, two halos collapse nearly together, and the one that forms stars first irradiates the other \citep{Dijkstra_2008, Visbal_2014_pair, Regan_2017}. \cite{rinaldi_not_2025} presented the first statistical census of LRD UV morphologies in the JADES fields and found that at least $30\%$ of LRDs have close companions in the rest-frame UV, challenging the initially assumed compact morphology inferred from rest-frame optical observations. Recently, \cite{Baggen_2026} and \cite{Barger_2026} reported that compact, UV-bright companions of LRDs lie at projected separations of $0.5-5$ kpc; \cite{Baggen_2026_resolving} and \cite{Golubchik_2026} also reported a similar case behind a lensing cluster.

However, the presence of a neighbor does not necessarily imply causality, because standard galaxies also cluster. Therefore, \cite{Pacucci_Urry_2026} measured companion statistics for $253$ LRDs against $>7000$ control galaxies drawn from the same mosaics, matched in redshift, brightness, and compactness. The brightest LRDs host blue companions within $2$ kpc at a rate exceeding that of the control sample, significant at $3.3\sigma$. Those companions are also overwhelmingly blue, and significantly bluer than the companions of the control galaxies. Hence, that study concluded that the association is physical.

All results mentioned above are based on photometry. Broad-band colors identify a blue, compact neighbor, but they cannot answer the four questions on which the interpretation is based: (i) whether the neighbor lies at the redshift of the LRD, (ii) whether its blue light is starlight, or rather nebular continuum that the LRD itself could power \citep{Chen_2025}, or even part of a larger host in which the nucleus sits \citep{Rinaldi_2025}, (iii) how massive the black hole is and how much it has grown since seeding, and (iv) whether the two objects are gravitationally bound. Sections \ref{subsec:association} to \ref{subsec:dynamics} answer them in this order.

In this Letter, we present the only two LRD--companion pairs in the literature with spectroscopy for both members. In ``Waltz'' the LRD GS-206858 and its companion GS-206860 sit at $z = 3.936$, separated by $3.57$ kpc in projection; in ``Tango'' the LRD GS-13971 and its companion sit at $z = 5.4816$, separated by $2.54$ kpc.
Both companions were first identified photometrically by \cite{Baggen_2026}, who estimated Lyman-Werner intensities at the LRD positions comparable to or above the critical value for direct collapse. We name the systems after the two dances, the closer pair taking the closer dance.

We describe the observations and the measurements in \S \ref{sec:data_methods}, and present our results in \S \ref{sec:results}. In \S \ref{sec:summary_concl}, we discuss what these systems imply for the nature of LRDs. Throughout, we adopt the Planck18 cosmology \citep{Planck_2018}, where $1''$ corresponds to $7.15$ and $6.13$ proper kpc at $z \approx 3.9$ and $z \approx 5.5$, respectively.

\section{Data and Methods}
\label{sec:data_methods}

This Section describes the data available for the two systems and the measurements we derive from them. Both pairs lie in GOODS-S and were observed with NIRSpec, one through the micro-shutter assembly (MSA) and the other with the integral field unit (IFU).

\subsection{Data}
\label{subsec:data}

``Waltz'' is a pair of JADES sources \citep{Rieke_2023_JADES, JADES_2025, JADES_2023}. We use the NIRSpec \citep{NIRSPEC_2022} MSA spectra of its two members from the public JADES DR3 release, retrieved from MAST: for the LRD GS-206858, the prism and the G140M, G235M, and G395M gratings from PID 1180 (observations o026 and o028, $\approx 3.1$ ks per grating per observation); for the companion GS-206860, the same four dispersers in PID 1286 and, independently, in PID 1180. The two programs placed their shutters differently (position angles of $207\arcdeg$ and $195\arcdeg$, see also Fig. \ref{fig:gallery}). The companion consists of two knots, A and B (\S\ref{sec:decomposition}): the PID 1286 slitlet captures the larger fraction of the light of knot A, the line emitter, but also much of the continuum of knot B, whereas the PID 1180 slitlet captures less light but is dominated by knot A. We adopt PID 1286 for all luminosities. The standard Level-3 pipeline products of PID 1286 (also from MAST) add a G395H spectrum of the companion at $R \approx 2700$.
The JWST data analyzed here are available on MAST: the JADES products at \dataset[doi:10.17909/8tdj-8n28]{https://doi.org/10.17909/8tdj-8n28}, and the pipeline products of PIDs 1180, 1286, and 5664 at \dataset[doi:10.17909/26rs-hg30]{https://doi.org/10.17909/26rs-hg30}.

``Tango'' pairs the LRD GS-13971, first identified as such by \cite{Matthee_2023, Matthee_2026} and \cite{Ishikawa_2026}, with a blue companion $0\farcs415$ away. The companion is the bright [\ion{O}{3}] clump at the south-eastern end of the kpc-scale structure that crosses this LRD (LRD-204851) in the NIRCam line maps of \cite{Ji_2026a}, and a prism spectrum presented by \cite{Ji_2026b} places it at the redshift of the LRD. We obtain its spectroscopy from the NIRSpec IFU under program GO 5664 (PI: J. Matthee). We use the Level-3 cubes in PRISM/CLEAR ($0.60-5.30$ $\mu$m, $6.4$ ks) and G395H/F290LP ($2.87-5.27$ $\mu$m, $36.2$ ks), both sampled with $0\farcs1$ spaxels through an eight-point cycling dither; the region at full depth spans $2\farcs8 \times 2\farcs8$, which entirely contains the pair.

For Tango, we use the 14 NIRCam bands from F090W to F480M, and five ACS and four WFC3 bands from the DAWN JWST Archive mosaics ($0\farcs05$ per pixel, see \citealt{Valentino_2023}) reduced with \textsc{grizli} \citep{Brammer_2023}. For Waltz, we use the same 14 NIRCam bands from the JADES mosaics ($0\farcs03$ per pixel, see \citealt{Rieke_2023, JADES_2023}).

\subsection{Separating the Pairs}
\label{sec:decomposition}

In Tango, the two sources are separated by only twice the PSF core width, so we must decompose the cube before any measurement. After subtracting a master background built from source-free spaxels, we construct the PSF empirically from the broad H$\alpha$ wings of the LRD ($400 < |v| < 1200 \, \rm km \, s^{-1}$), which trace the unresolved broad-line region alone, following the established approach for separating quasars from their hosts in IFU data \citep{Husemann_2013, Marshall_2023, Decarli_2024}. The PSF requires a core of FWHM $0\farcs181$ and a halo of FWHM $0\farcs507$ carrying $\approx 39 \%$ of the flux. We then fit each wavelength plane within $0\farcs80$ of the pair midpoint with templates for the LRD, the companion, a third field source (a foreground galaxy, \S\ref{sec:photometry_orbits}), and a quadratic background, rescaling the errors to $\chi^2_\nu = 1$ on line-free planes; the template amplitudes as a function of wavelength are the extracted spectra. The two main templates are nearly orthogonal, and the companion receives only $\approx 3.4\%$ of the LRD's broad-line flux.
Two checks calibrate the decomposition: first, $20 \pm 6\%$ of the LRD's wing flux ends up in the background term, so the companion's template fluxes are lower limits at that level; and second, the F430M excess of the LRD in the NIRCam image, obtained without the cube, matches the total H$\alpha$ flux of the extracted spectra.

Waltz requires no such treatment, since its two members were observed as separate MSA targets. The structure to be resolved lies inside the companion, which the imaging splits into two knots $0\farcs23$ apart. In the 2D prism spectrum, the emission lines and the UV continuum trace knot A, while the continuum beyond $2.5 \, \mu$m is displaced by $0\farcs050$ toward knot B. We cannot cleanly separate the two knots at the $0\farcs24$ effective spatial resolution along the slit. Hence, every spectroscopic quantity we quote for Waltz's companion refers to knot A.

\subsection{Emission Lines and Velocity Offsets}
\label{sec:lines}

We fit every line as part of a complex, with a local continuum, a single redshift and width shared by the narrow components, and the [\ion{N}{2}] doublet fixed at its theoretical ratio of $2.94$ \citep{Osterbrock_Ferland_2006}. All fits use standard least-squares methods from \textsc{astropy} and \textsc{scipy} \citep{Scipy_2020, Astropy_2022}, with uncertainties from a parametric bootstrap ($300$ refits), and the error arrays rescaled so that $\chi^2_\nu = 1$ away from the lines. Intrinsic widths require care, because the spectral resolution of an MSA spectrum depends on how the source fills the shutter. Therefore, we quote $\sigma_{\rm int}$ between two bracketing line-spread functions, for a uniformly illuminated shutter and for a compact source; the [\ion{O}{3}] line of the Waltz LRD, for instance, is unresolved under the first and gives $52 \, \rm km \, s^{-1}$ under the second.

\begin{deluxetable*}{lcc}
\tablecaption{Properties of the two LRD--companion pairs.\label{tab:properties}}
\tablehead{
  \colhead{} &
  \colhead{\textcolor{waltzcol}{\textbf{Waltz}}} &
  \colhead{\textcolor{tangocol}{\textbf{Tango}}} \\[-2pt]
  \colhead{} &
  \colhead{\textcolor{waltzcol}{$z = 3.936$}} &
  \colhead{\textcolor{tangocol}{$z = 5.4816$}}
}
\startdata
\multicolumn{3}{c}{\secrow{LRD--Companion Pairs}} \\
\rowlab{Velocity offset, $\Delta v$ (companion $-$ LRD)} & $+131.6 \pm 6$ km s$^{-1}$ & $-56.8 \pm 5$ km s$^{-1}$ \\
\rowlab{Projected separation} & $3.57 \pm 0.03$ kpc & $2.54 \pm 0.15$ kpc \\
\rowlab{Bound fraction, $P_{\rm bound}$} & $0.48$ ($0.17$--$0.71$) & $0.81$ ($0.71$--$0.90$) \\
\rowlab{Time to merger} & $457^{+630}_{-243}$ Myr & $129^{+352}_{-96}$ Myr \\
\rowlab{Minimum mass for a bound orbit, $M_{\rm bind}$} & $> 7.2 \times 10^{9}\Msun$ & $> 9.5 \times 10^{8}\Msun$ \\
\multicolumn{3}{c}{\secrow{Little Red Dots}} \\
\rowlab{Broad H$\alpha$ FWHM} & $1169^{+108}_{-88}$ km s$^{-1}$ & $2132 \pm 18$ km s$^{-1}$ \\
\rowlab{Broad H$\alpha$ profile} & unabsorbed & absorbed core \\
\rowlab{Black hole mass (virial), $\log (\Mblack/\Msun)$} & 6.3--6.6 & 7.5--7.8 \\
\rowlab{Bolometric luminosity, $L_{\rm bol}$} & $1.4 \times 10^{44}$ erg s$^{-1}$ & $1.2 \times 10^{45}$ erg s$^{-1}$ \\
\rowlab{Eddington ratio, $\fedd$} & 0.55 & 0.34 \\
\rowlab{Narrow-line dust attenuation, $A_V$} & $\simeq 0$ & $\simeq 0.9$ mag \\
\rowlab{Narrow-line velocity dispersion, $\sigma$} & $\leq 52$ km s$^{-1}$ & 32--54 km s$^{-1}$ \\
\rowlab{Host stellar mass, $\Mstar$} & {$< 3.3 \times 10^{9}$}$\Msun$ & $< 8.3 \times 10^{9}\Msun$ \\
\multicolumn{3}{c}{\secrow{Blue Companions}} \\
\rowlab{H$\alpha$ rest-frame equivalent width} & 800--950 \AA & $733 \pm 80$ \AA \\
\rowlab{Dust attenuation, $A_V$ (Balmer decrement)} & $0.4 \pm 0.4$ mag & $\simeq 0$ \\
\rowlab{Gas-phase metallicity} & $\simeq 0.2\ Z_\odot$ & $\simeq 0.2\ Z_\odot$ \\
\rowlab{Stellar mass, $\Mstar$} & $6.6 \times 10^{7}\Msun$ & $1.7 \times 10^{8}\Msun$ \\
\rowlab{Star formation rate (H$\alpha$)} & $1.2\Msun$ yr$^{-1}$ & $2.6\Msun$ yr$^{-1}$ \\
\rowlab{Growth timescale, $\Mstar / {\rm SFR}$} & 56 Myr & 65 Myr \\
\rowlab{Velocity dispersion, $\sigma$} & 20--50 km s$^{-1}$ & 47--65 km s$^{-1}$
\enddata
\tablecomments{Values for the Waltz companion refer to knot A, deblended from knot B in the photometry (\S\ref{sec:photometry_orbits}); the MSA spectra of knot A contain some continuum of knot B (\S\ref{sec:decomposition}). $M_{\rm bind} = r_p \Delta v^2/(2G)$ is the mass that must lie within the projected separation for the pair to be bound at the observed line-of-sight velocity; it is a lower limit, since the true separation and speed exceed the projected ones. $P_{\rm bound}$ is marginalized over the halo-mass prior of \S\ref{subsec:dynamics}, and the range in parentheses spans the bound fraction across that prior (Fig.~\ref{fig:dynamics}a); merger times are medians with 16th--84th percentiles of the bound realizations. Black hole masses are the range across the calibrations of \citet{Greene_2005} and \citet{Reines_2013}; single-epoch estimates carry an additional systematic scatter of $0.3-0.5$ dex. $L_{\rm bol}$ and $\fedd$ follow from the broad H$\alpha$ luminosity and carry a systematic uncertainty of $\approx 0.3$ dex from the bolometric correction, which $\fedd$ inherits together with that of $\Mblack$. Systematic uncertainties are $\approx 0.1$ dex on the gas-phase metallicities (strong-line calibration), $0.3$ dex on the stellar masses (templates), and $0.1$ dex on the star formation rates (metallicity of the conversion). Host stellar masses are upper limits from stars-only fits, dust-free for Waltz (\S\ref{sec:photometry_orbits}).}
\end{deluxetable*}

We measure velocity offsets as companion minus LRD from [\ion{O}{3}]$\lambda 5008$, the strongest line in all four sources.
Both companions lie at the redshift of their LRD: the Waltz LRD and companion are at $z = 3.9363$ and $3.9386$, the Tango LRD and companion at $z = 5.4816$ and $5.4804$. Throughout, we label each system by the [\ion{O}{3}] redshift of its LRD.
For Waltz, the two MSA observations of the companion give $+134.0 \pm 4.2$ and $+127.6 \pm 5.5 \, \rm km \, s^{-1}$ against the same LRD spectrum, and we adopt their weighted mean (their difference is a systematic term). For Tango, the offset is measured on the decomposed G395H spectra, and the systematic is half the spread across the template variants of \S \ref{sec:decomposition}; the companion's own [\ion{O}{3}] and H$\alpha$ velocities agree to $0.4 \, \rm km \, s^{-1}$.

The same line fitter provides the broad H$\alpha$ components. For Waltz, we fit one broad Gaussian shared across three independent extractions of the G395M data. For Tango, we fit the decomposed G395H spectrum.

\subsection{Photometry and SED Fitting}
\label{sec:photometry_orbits}

We measure photometry identically in both systems. We convolve all bands to the F480M resolution (FWHM $0\farcs160$), and extract fluxes in apertures of radius $0\farcs25$ for the LRDs and $0\farcs15$ for the companions, using a local background annulus. We separate blended components by fitting Gaussian components of the matched PSF width simultaneously at the source positions; for the Waltz companion this assigns the flux of knots A and B in every band, and all quantities quoted for the companion refer to knot A. Total fluxes follow each LRD's growth curve in the long-wavelength bands, with an encircled energy of $0.86 \pm 0.05$ within $0\farcs30$. The two knots of the Waltz companion are $0\farcs23$ apart, so residual cross-contamination from the PSF wings remains possible, and we include it in the quoted systematic uncertainty.

We checked our totals against the DJA catalog \citep{Valentino_2023}. For the Waltz LRD, the two measurements agree within $2\%$, while for the Tango LRD within $10\%$. For the companions, the catalog totals are $1.8\times$ and $2.9\times$ ours, so the pairs require a deblended, PSF-matched scheme.

We derive stellar properties via SED fitting. We first obtain stellar masses from \textsc{EAZY} (version 0.8.7, \citealt{Brammer_2008}) with its 13 redshift-dependent \texttt{corr\_sfhz\_13} templates, which include nebular emission, and a Chabrier initial mass function \citep{Chabrier_2003}. We run each fit twice. The first run leaves the redshift free and serves two purposes. It places the sources without spectra: knot B in Waltz at $z_{\rm phot} = 3.7$ ($3.5-4.0$), consistent with the pair; the source within $0\farcs8$ of the Tango pair at $z_{\rm phot} = 2.8$ ($2.7-3.0$), with flux detected below the Lyman limit, i.e., a foreground galaxy; and a second blue source $8.6$ kpc from the Tango LRD at $z_{\rm phot} = 5.6$ ($5.4-5.7$). For the companions, the Tango one is placed at $z_{\rm phot} = 5.6$ ($5.4-5.7$), whereas the strong-line SED of the Waltz companion is degenerate between $z \approx 0.7$ and $z \approx 3.9$.

The second run fixes the redshift to the pair redshift and yields the masses. For the companions, template masses carry systematic uncertainties of $\approx 0.3$ dex. For the LRDs, the same procedure assumes all the light is stellar, leading to an upper limit in the sense that stars cannot contribute more light than observed. The templates reach the red rest-optical continuum only with heavy attenuation, $A_V = 1.4$ mag for Waltz (an old population, with $\Mstar/{\rm SFR} \approx 2$ Gyr) and $A_V = 2.6$ mag for Tango (a young one, with ${\rm SFR} \approx 100 \Msun \, {\rm yr}^{-1}$ and $\Mstar/{\rm SFR} \approx 90$ Myr). We do not regard this attenuation as physical for Waltz, whose narrow lines show none (\S\ref{subsec:black-holes}). We therefore repeat the fits with the dust-free templates alone (Appendix~\ref{app:sed}). The Waltz fit then gives $3 \times 10^{9} \Msun$, which we adopt as its limit. Tango, whose narrow lines do show dust, changes little, from $8 \times 10^{9}$ to $7 \times 10^{9} \Msun$, and we keep the former. Both limits are consistent with recent literature (e.g., \citealt{Chen_2025b}). We fit the photometry rather than the spectra because the stellar mass needs total fluxes, which the MSA spectra (slit losses, part of knot B in the slitlet) and the decomposed IFU spectra (template fluxes that are lower limits) do not provide; the spectra constrain the lines, from which we take metallicity, dust, and star formation rate.

As an independent check on the companions, we also used \textsc{bagpipes} \citep{carnall_vandels_2019}, with templates from  \citet{Bruzual_Charlot_2003}, a Kroupa IMF, nebular emission from \textsc{Cloudy} \citep{Cloudy}, a non-parametric continuity star formation history \citep{leja_how_2019}, and broad priors on mass, attenuation, and metallicity. The \textsc{bagpipes} masses of the two companions, $\log(\Mstar/\Msun) = 8.0$ and $8.3$, agree with the \textsc{EAZY} values, $7.8$ and $8.2$, within the $0.3$ dex template systematic.

Both codes yield low stellar metallicities ($\lesssim 30\%\,Z_{\odot}$); for Waltz the value refers to knot A. The gas-phase metallicity measured from the emission lines is $\approx 0.2\,Z_\odot$ for both companions (\S\ref{subsec:companions}); we adopt it in the H$\alpha$-to-SFR conversion below, with $0.1-0.3\,Z_\odot$ as the systematic range.

In both Tango and Waltz, H$\alpha$ shows large equivalent widths, indicating that the companions are actively forming stars. With the metallicity-dependent conversions, the ratio $\rm SFR(H\alpha)/SFR(UV)$ is $\approx 1.5$ for the Waltz companion and consistent with unity for the Tango companion, whose template H$\alpha$ flux is a lower limit (\S\ref{sec:decomposition}); a ratio above unity indicates a rising star formation rate, as commonly observed at such low stellar masses (e.g., \citealt{navarro-carrera_burstiness_2026}). The absolute values depend on the adopted metallicity at the $0.1$ dex level \citep{theios_dust_2019}. Similarly, their inferred star-formation histories from \textsc{bagpipes} show a strong upturn over the last 100 Myr. We discuss the implications in \S \ref{subsec:companions}.

\section{Results}
\label{sec:results}

We present the two pairs, Tango and Waltz, in parallel throughout this Section. Table \ref{tab:properties} collects all measured quantities and their uncertainties for both systems, while the main text discusses only the relevant numbers.

\subsection{Two Physically Associated Pairs}
\label{subsec:association}

\begin{figure*}[!tp]
\centering
\includegraphics[width=\textwidth]{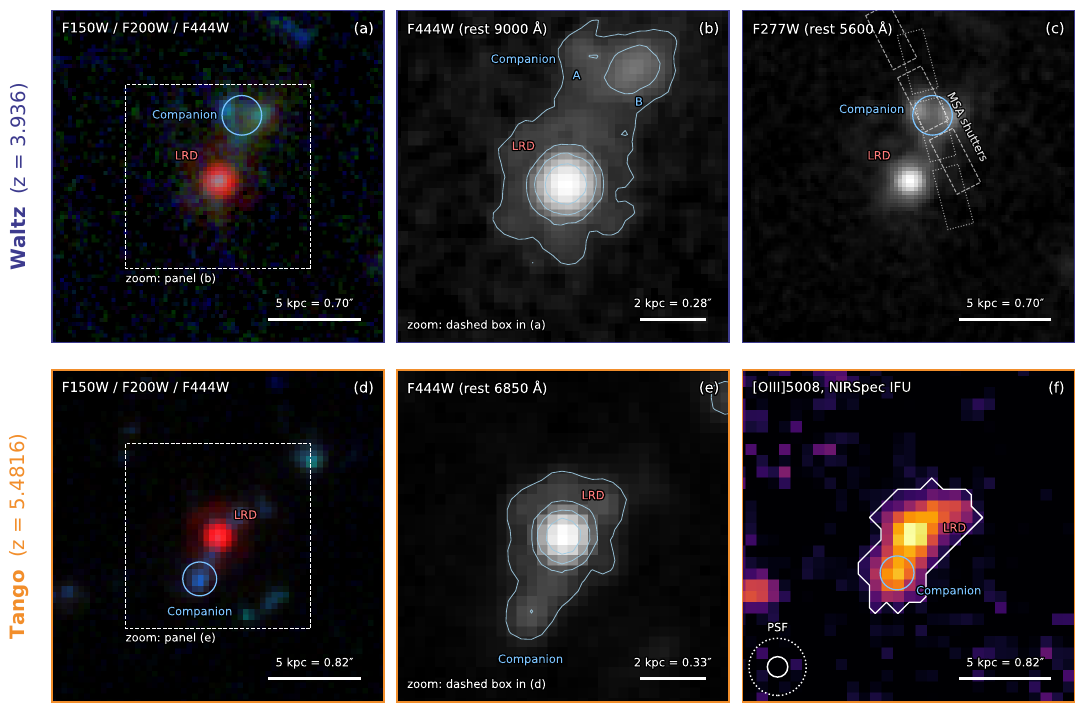}
\caption{The two LRD--companion pairs. Top row: Waltz ($z = 3.936$). Bottom row: Tango ($z = 5.4816$). \textit{(a), (d):} NIRCam composites of F150W, F200W, and F444W (blue, green, and red channels); the dashed box marks the region shown in the adjacent panel. \textit{(b), (e):} F444W alone at $1.8\times$ zoom ($10 \times 10$~kpc, the dashed box of the adjacent $18 \times 18$~kpc panel), sampling the rest-frame optical, where the LRD dominates; the zoom resolves the two knots of the Waltz companion (A and B, $0\farcs23$ apart) and shows the compactness of both LRDs at this resolution. \textit{(c):} F277W, which contains [\ion{O}{3}] and H$\beta$ at $z = 3.936$, where the companion is conspicuous; the dashed and dotted rectangles are the open MSA shutters of the two companion observations. \textit{(f):} [\ion{O}{3}]$\lambda 5008$ map from the NIRSpec IFU, integrated around the pair's systemic velocity; the white outline encloses the region where the line is detected, and the companion appears as a distinct peak within it. The projected separations are $3.57$ and $2.54$ kpc. The concentric circles in the lower left of (f) show the FWHM of the core ($0\farcs18$, solid) and halo ($0\farcs51$, dotted) of the empirical IFU PSF (\S\ref{sec:decomposition}).}
\label{fig:gallery}
\end{figure*}

Every claim in this Letter is based on one fact: the two blue sources are neighbors of their LRDs. Figure \ref{fig:gallery} shows the two systems in the configuration that photometric surveys have reported around bright LRDs \citep{Baggen_2026, Barger_2026, Pacucci_Urry_2026} but could not spectroscopically confirm, except for a prism redshift of the Tango companion \citep{Ji_2026b}. The projected separations are $3.57 \pm 0.03$ kpc for Waltz and $2.54 \pm 0.15$ kpc for Tango.

\begin{figure*}[!tp]
\centering
\includegraphics[width=\textwidth]{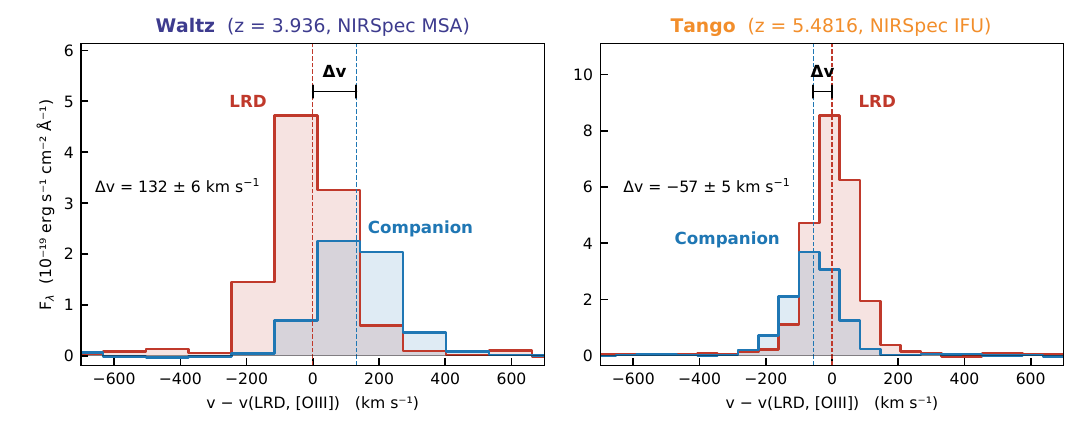}
\caption{Velocity offsets of the two pairs. [\ion{O}{3}]$\lambda 5008$ profiles of each LRD (red) and its companion (blue), in velocity relative to the LRD centroid; dashed lines mark the two centroids. Left: Waltz, from the JADES G235M spectra, $\Delta v = +131.6 \pm 3.3 \, \rm km \, s^{-1}$ (weighted mean of two independent MSA observations of the companion). Right: Tango, from the G395H cube, $\Delta v = -56.8 \pm 0.7 \, \rm km \, s^{-1}$. Systematic terms are included in Table~\ref{tab:properties}.}
\label{fig:kinematics}
\end{figure*}

Both companions lie at the redshift of their LRD (Fig. \ref{fig:kinematics}). In Waltz, [\ion{O}{3}]$\lambda 5008$ places the companion at $\Delta v = +131.6 \pm 6 \, \rm km \, s^{-1}$, and the two MSA configurations agree within their errors. In Tango, the decomposed cube gives $\Delta v = -56.8 \pm 5 \, \rm km \, s^{-1}$, and the companion's own [\ion{O}{3}] and H$\alpha$ velocities agree to $0.4 \, \rm km \, s^{-1}$. Each companion shows H$\beta$, [\ion{O}{3}], and H$\alpha$ at a common redshift (Fig. \ref{fig:spectra}).

Both velocity offsets are within $\vert \Delta v \vert < 500 \, \rm km \, s^{-1}$, the conventional limit for two galaxies to belong to the same structure at these redshifts \citep{Ventou_2017, Helton_2024}. For Tango, the orbital ensemble of \S \ref{subsec:dynamics} places the companion at a three-dimensional speed of $\approx 114 \, \rm km \, s^{-1}$, against a median escape speed of $315 \, \rm km \, s^{-1}$ at its position.  Two systems do not make a population, but they show that the photometric selection of \cite{Baggen_2026} finds real pairs.

Neither system is a clean pair. In Waltz, the companion consists of two knots $0\farcs23$ ($1.6$ kpc) apart: A, which emits the lines and the UV continuum, and B, which shows a Balmer break and a stellar mass of $4 \times 10^{8} \Msun$ (\S\ref{subsec:companions}), with faint emission around both. In Tango, the source within $0\farcs8$ of the pair is a foreground galaxy ($z_{\rm phot} = 2.8$; \S\ref{sec:photometry_orbits}), but a second blue source $1\farcs4$ ($8.6$ kpc) from the LRD has $z_{\rm phot} = 5.6$ and $8 \times 10^{7} \Msun$ of stars, and the [\ion{O}{3}] map (Fig.~\ref{fig:gallery}f) shows line emission between the LRD and the companion. Diffuse emission between the members is what an interacting pair should show.

\subsection{Two Similar Companions Half a Gyr Apart}
\label{subsec:companions}

\begin{figure*}[!tp]
\centering
\includegraphics[width=\textwidth]{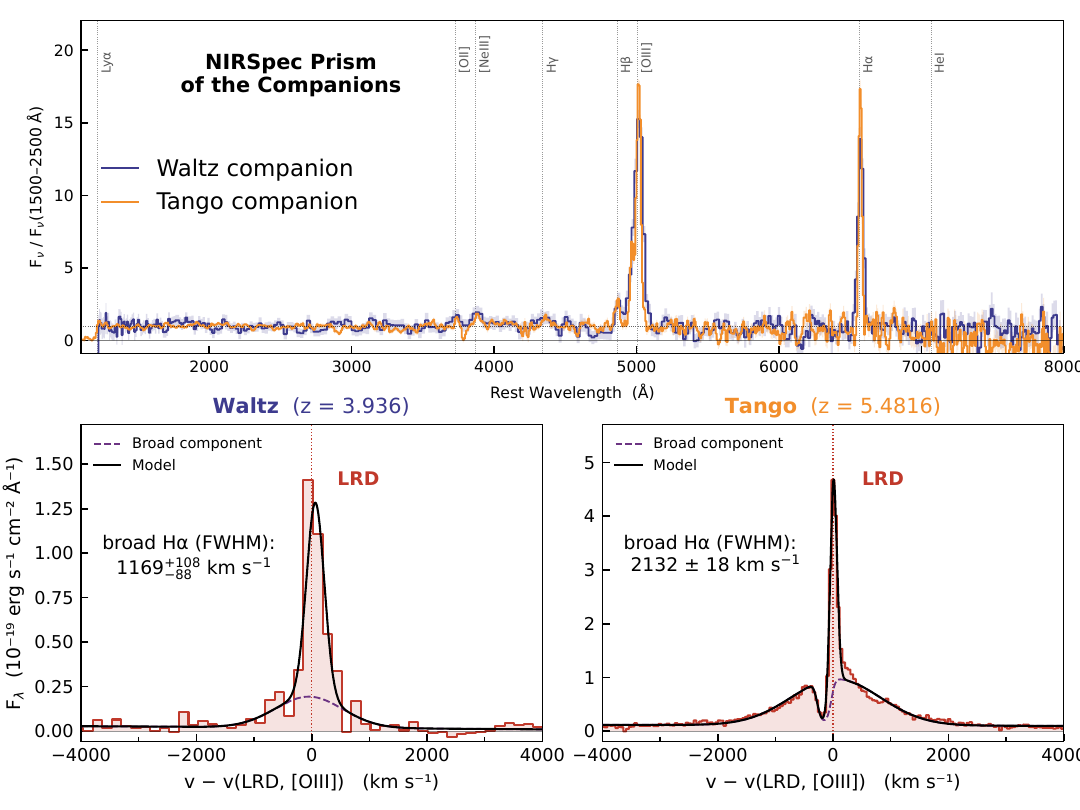}
\caption{Spectra of the companions and of the LRDs. Top: NIRSpec prism spectra of the two companions, normalized to their mean flux density at $1500-2500$ \AA\ rest-frame; shading is the $1\sigma$ uncertainty and dotted lines mark the principal transitions. The two are nearly indistinguishable across $524$ Myr of cosmic time: blue continua, little or no dust attenuation, and extreme equivalent widths, EW$_0$(H$\alpha$) $= 733 \pm 80$ \AA\ for Tango and $800-950$ \AA\ for Waltz. Bottom: H$\alpha$ complexes of the two LRDs, in velocity relative to their [\ion{O}{3}] centroids, with the best-fit model (solid; narrow H$\alpha$, the [\ion{N}{2}] doublet, and a broad Gaussian over a linear continuum) and its broad component (dashed) overplotted. Waltz shows a clean broad line of FWHM $1169^{+108}_{-88} \, \rm km \, s^{-1}$. Tango shows a broad line of FWHM $2132 \pm 18 \, \rm km \, s^{-1}$ detected at $109\sigma$, whose core is absorbed: the dip in its broad component is the Gaussian absorber of \S\ref{subsec:black-holes}, blueshifted by $\approx 155 \, \rm km \, s^{-1}$.}
\label{fig:spectra}
\end{figure*}

The second question is what the blue light is. Figure \ref{fig:spectra} (top) shows a first finding: normalized in the rest-UV, the prism spectra of the two companions are indistinguishable, from the Lyman break to H$\alpha$ and beyond. They share the same flat blue continuum and emission lines with the same relative strengths. Two galaxies separated by $524$ Myr of cosmic time, selected only for sitting beside an LRD, have spectra that cannot be told apart at this resolution: each is a young starburst caught within its first tens of Myr. 

The H$\alpha$ equivalent widths, $800-950$ \AA\ in Waltz and $733 \pm 80$ \AA\ in Tango, are reached only by a single burst younger than $\approx 6$ Myr, or by continuous star formation younger than $\approx 100$ Myr \citep{Leitherer_1999, Eldridge_2017}. The gas that formed the stars is metal-poor \citep{Curti_2020}: $12 + \log({\rm O/H}) \simeq 8.0$, i.e., $0.2 \, Z_\odot$ in both companions. The same finding is confirmed by the [\ion{N}{2}]/H$\alpha$ ratios, $< 0.06$ for the Waltz companion (G395H) and $0.053 \pm 0.010$ for the Tango companion, measured on the decomposed G395H spectrum ($R \approx 2700$), where
[\ion{N}{2}]$\lambda 6583$ is $\sim 900$ km s$^{-1}$ from H$\alpha$ and fully resolved. The gas carries little dust. The Balmer decrement of the Tango companion, $2.53 \pm 0.12$ (from its G395H spectrum, where H$\beta$ and H$\alpha$ are resolved from [\ion{O}{3}] and [\ion{N}{2}]), lies at or slightly below the Case B value ($2.86$ at $10^{4}$ K, $2.75$ at $2 \times 10^{4}$ K, as expected for metal-poor gas), so its gas carries no measurable dust. The two grating observations of the Waltz companion give $2.8 \pm 0.4$ and $3.8 \pm 0.5$, with a weighted mean of $3.2 \pm 0.3$, i.e., $A_V = 0.4 \pm 0.4$ mag for a Calzetti law \citep{Calzetti_2000}; the Waltz companion is dust-poor rather than dust-free. The gas is cold, with intrinsic dispersions of $20-50$ and $47-65 \, \rm km \, s^{-1}$. 

The SED fits of \S\ref{sec:photometry_orbits} give stellar masses of $6.6 \times 10^{7}$ and $1.7 \times 10^{8} \Msun$, against star formation rates of $1.2$ and $2.6 \Msun \, {\rm yr}^{-1}$, from the metallicity-dependent H$\alpha$ conversion of \citet{KorhonenCuestas_2025} at the measured $Z \simeq 0.2\,Z_\odot$ (adopting $0.1$ or $0.3\,Z_\odot$ changes them by $\pm 0.05$ dex): both galaxies assembled most of their stars within the last $100$ Myr; at these rates, the SED masses build up in $56$ and $65$ Myr (Table~\ref{tab:properties}).
Non-parametric star formation histories fitted with \textsc{bagpipes} agree: in both companions the star formation rate rises steeply over the last $\sim 100$ Myr to its present H$\alpha$ value, while older stars are only loosely constrained. In Waltz, older stars do sit next to the burst: knot B, $0\farcs23$ from the line-emitting knot A, shows a Balmer break between F150W and F182M and has a stellar mass of $4 \times 10^{8}\Msun$.

This light is starlight from a distinct stellar system, not nebular gas photoionized by the LRD, which is the first alternative that broad-band colors left open \citep{Chen_2025b, Chen_2025}, for three reasons. First, the velocities, as each companion moves at its own systemic velocity, $132$ and $57 \, \rm km \, s^{-1}$ away from its LRD, and its lines are narrow; gas merely illuminated by the LRD would move with the LRD. Second, the line ratios match star-forming gas. Third, each companion is a compact source with its own stellar continuum. Whether that stellar system is a separate galaxy or a star-forming clump within a common structure \citep{Rinaldi_2025} is a different question, which these data do not settle: the diffuse emission between the two members (Fig.~\ref{fig:gallery}b, e, f) suggests that they are connected, as an interacting pair should be.

In the rest-frame UV, the companions are as luminous as their LRDs. Tango's companion equals its LRD in F115W and outshines the compact nucleus alone by a factor of two; the same holds for Waltz's companion when its two knots are counted together. In summary, each LRD sits a few kpc from a bright source of ultraviolet radiation, observed within tens of Myr of a burst.

\subsection{A Black Hole in the Heavy-Seed Mass Range}
\label{subsec:black-holes}

The third question concerns the LRDs themselves: how massive are their black holes? Waltz gives the cleaner answer of the two, because its broad H$\alpha$ line (Fig. \ref{fig:spectra}, bottom left) is a single unabsorbed component. Its FWHM is $1169^{+108}_{-88} \, \rm km \, s^{-1}$, and the broad component is required by the data at $\Delta \chi^2 = 284$. Virial calibrations bracket the mass between $\log (\Mblack/\Msun) = 6.31$ and $6.59$ \citep{Greene_2005, Reines_2013}, that is, $2-4 \times 10^{6} \Msun$, radiating at $\fedd = 0.55$ with $L_{\rm bol} = 1.4 \times 10^{44} \, \rm erg \, s^{-1}$.

Direct collapse produces black holes of $\sim 10^{4}-10^{6} \Msun$ \citep{Ferrara_2014}, so Waltz's black hole is only a factor of a few above that range. At its measured Eddington ratio, with a radiative efficiency of $0.1$, the growth from a $10^6 \Msun$ seed, the top of the direct collapse range, to the observed $2-4 \times 10^6 \Msun$  takes $60-130$ Myr, about one e-folding, or longer if accretion was interrupted.

The gas around the black hole is as metal-poor as the companion's, and in Waltz it carries no dust. The narrow-line regions of both LRDs sit at the saturated maximum of the strong-line calibration, with [\ion{O}{3}]/H$\beta = 8.1 \pm 1.2$ in Waltz and $7.4$ in Tango, and with [\ion{N}{2}]/H$\alpha < 0.08$ and $0.008$. Waltz's narrow Balmer decrement is at the Case B value, so its narrow-line gas carries no dust, while Tango's implies $A_V \simeq 0.9$ mag. Both narrow-line regions are as cold as the companions, with dispersions of $\leq 52$ and $32-54 \, \rm km \, s^{-1}$, and the hosts are limited to $\Mstar < 3 \times 10^{9}$ and $8 \times 10^{9} \Msun$. In both systems, the red light is dominated by a compact source (Fig.~\ref{fig:gallery}b, e) while the rest-UV light is extended, so the host is seen mainly in the ultraviolet. 

In Waltz, a black hole of a few $10^{6} \Msun$ accretes at half the Eddington rate in metal-poor, dust-free gas, in a host with $\Mstar < 3 \times 10^{9} \Msun$, alongside a starburst of the same gas. Every element of the environment predicted to form a heavy seed is present.

Tango hosts a more evolved black hole and its mass is less certain. Its broad H$\alpha$ (Fig. \ref{fig:spectra}, bottom right) is detected at $109\sigma$ with FWHM $2132 \pm 18 \, \rm km \, s^{-1}$, matched by a broad H$\beta$ of $2390 \pm 252 \, \rm km \, s^{-1}$, but its core is absorbed: a single Gaussian leaves a $-24\sigma$ residual within $300 \, \rm km \, s^{-1}$ of the line center. Modeling the absorber as a Gaussian in optical depth acting on the broad component (Fig.~\ref{fig:spectra}, bottom right) places it at $v_{\rm abs} \approx -155 \, \rm km \, s^{-1}$ relative to the [\ion{O}{3}] systemic velocity, with a FWHM of $\approx 180 \, \rm km \, s^{-1}$ and a peak optical depth of $\approx 2$: dense gas moving toward us along the line of sight, as in the other LRDs with Balmer absorption \citep{Deugenio_2025, juodzbalis_incomplete_2026}. The Waltz data show no such feature (Fig.~\ref{fig:spectra}, bottom left), but an absorber of this width would be unresolved and diluted at the G395M resolution, so its absence there is not constraining. 

With the core removed, both the width and the flux of the broad component depend on the profile assumed, leading to $\log (\Mblack/\Msun) = 7.46-7.80$, an order of magnitude above Waltz, at $\fedd = 0.34$ and $L_{\rm bol} = 1.2 \times 10^{45} \, \rm erg \, s^{-1}$. The two systems thus sample two stages of the same growth: one black hole about one e-folding past its birth mass and one several beyond it.
These are single-epoch virial masses: if electron scattering in a dense envelope broadens the lines \citep{Rusakov_2025, Modelevsky_2026}, the true masses are lower and the Eddington ratios correspondingly higher, since $L_{\rm bol}$ does not depend on the line width. A mass ten times lower would place the Waltz black hole at $\approx 3 \times 10^{5} \Msun$, inside the direct collapse birth range and radiating at $\fedd \approx 5$, with no growth required since seeding, and Tango at $\approx 4 \times 10^{6} \Msun$ with $\fedd \approx 3$. The physical association, the companion properties, and the orbital results do not depend on $\Mblack$, which is less than $0.1\%$ (Waltz) and $7\%$ (Tango) of the minimum binding mass. What changes, of course, is the interpretation, as lower masses make the black holes less overmassive relative to their hosts, and the heavy-seed inference then rests on the environment.

\begin{figure*}[!tp]
\centering
\includegraphics[width=\textwidth]{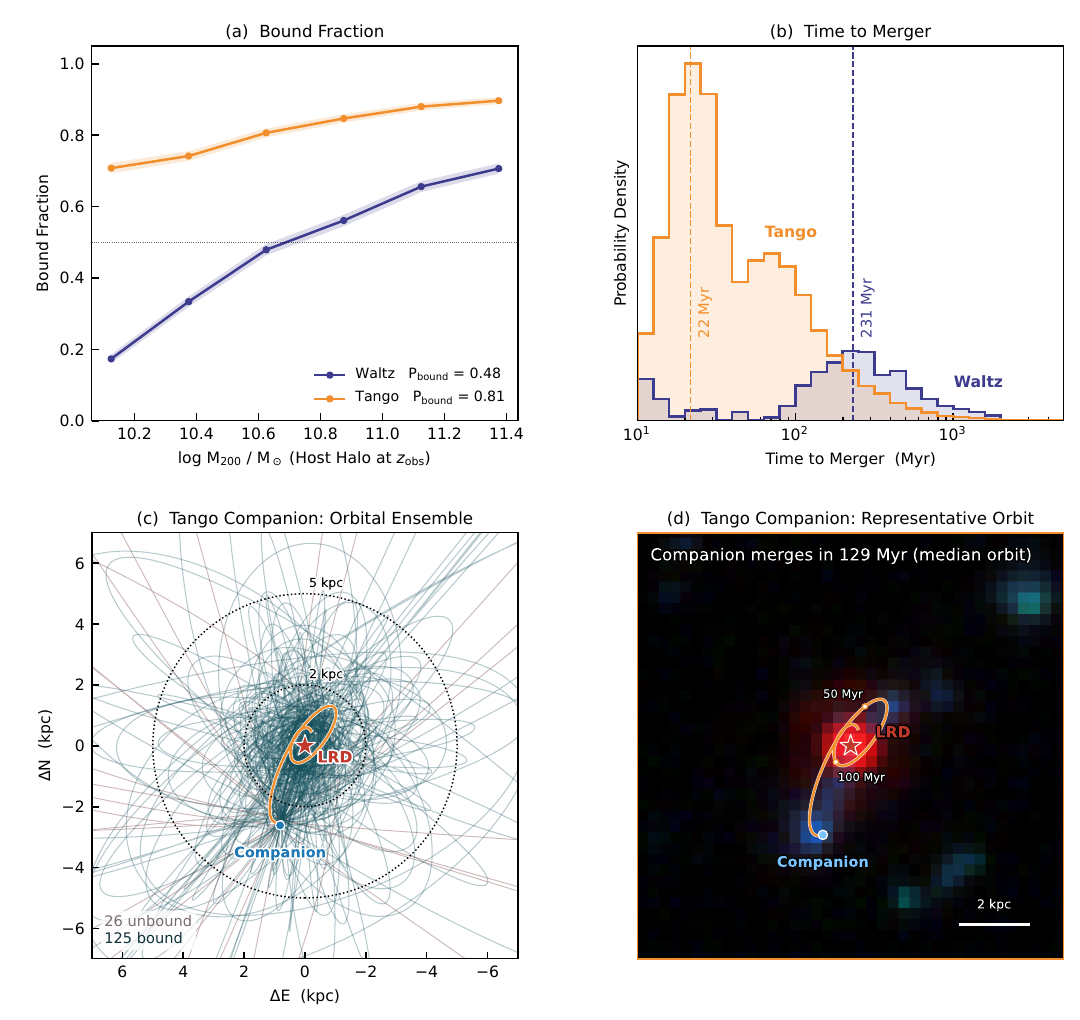}
\caption{Orbital analysis, from Monte Carlo ensembles of $6000$ orbits per system. \textit{(a):} Fraction of bound orbits against the host halo mass: the bound fraction spans $0.17$--$0.71$ for Waltz and $0.71$--$0.90$ for Tango across the prior ($0.48$ and $0.81$ marginalized). Halo masses refer to the epoch of observation; the $\sim 10^{8} \Msun$ atomic-cooling halo in which a heavy seed forms at $z \gtrsim 10$ grows into this range by $z = 4-6$ (\S\ref{sec:summary_concl}). \textit{(b):} Merger-time distributions of the bound orbits. Dashed lines mark the modes, $22$ Myr for Tango and $231$ Myr for Waltz (medians are $129$ Myr and $457$ Myr, respectively). \textit{(c):} A subset of the Tango ensemble ($125$ bound and $26$ unbound trajectories) with the median orbit highlighted. \textit{(d):} The median orbit, with $t_{\rm merge} = 129$ Myr, projected on the RGB composite, with marked positions after $50$ and $100$ Myr.}
\label{fig:dynamics}
\end{figure*}

\subsection{Two Pairs on Their Way to a Merger}
\label{subsec:dynamics}

The fourth question is whether the companions are bound to their LRDs, and what becomes of them. A spectroscopic pair offers two observables: (i) the projected separation and (ii) the line-of-sight velocity. As such, whether these describe a bound orbit depends on the mass between the two galaxies. 
We use a Monte Carlo ensemble of $6000$ orbital realizations per system to infer the orbits from $r_{\rm p}$ and $\Delta v$. The host is an NFW halo \citep{NFW_1996} of mass $M_{200}$ drawn log-uniformly over $10^{10}-10^{11.5} \Msun$, with a concentration drawn uniformly over $3-6$ (Waltz) or $2.5-5$ (Tango), plus a Hernquist baryonic component \citep{Hernquist_1990} with a scale length of $0.3$ kpc and a mass drawn log-uniformly from $10^8 \Msun$ to $8 \times 10^{9} \Msun$, so that the orbits do not depend on a specific host stellar mass. Physics bounds the halo prior on both sides. Below $10^{10} \Msun$, an NFW halo with concentration $3-6$ contains only $2-3 \times 10^{9} \Msun$ within $3.6$ kpc, under half of what Waltz requires, so lower masses would add only realizations that are unbound by construction. At the upper end, halos become rarer than LRDs: those above $3 \times 10^{11} \Msun$ have a number density of $\approx 10^{-4} \, \rm cMpc^{-3}$ at $z = 5.5$ \citep{Tinker_2008}, the upper end of the LRD density, while at $z = 3.9$ halos up to $10^{12} \Msun$ remain as common as LRDs \citep{Pacucci_Loeb_2025}, so the ceiling is conservative for Waltz and raising it would only increase its bound fraction (Fig.~\ref{fig:dynamics}a). The satellite is an NFW subhalo of concentration $5$ (Waltz) or $4$ (Tango), with a mass drawn log-uniformly from the stellar mass of the companion, $5 \times 10^{8} \Msun$ for knots A and B together and $1.7 \times 10^{8} \Msun$ for Tango, up to $10^{10} \Msun$ and at most half the host mass; the stars are never stripped.

Which object sits at the center of the potential is not known. We place the LRD at the center, because it hosts the black hole, dominates the rest-optical light, and coincides with the peak of the [\ion{O}{3}] emission (Fig.~\ref{fig:gallery}f); in the direct collapse picture, the seed may instead form in a satellite of the irradiating galaxy \citep{Agarwal_2013}. For Waltz, the data allow either configuration: the companion holds $5 \times 10^{8} \Msun$ of stars in its two knots, while the LRD host has only an upper limit. The velocity dispersions do not settle it. From its dispersion, the companion has a dynamical mass of $10^{8}-10^{9} \Msun$; Waltz needs more than that to be bound. Note that the bound fraction is the same whichever member is central, but the merger time is not: a heavier satellite sinks faster. In fact, the median drops from $630$ to $357$ Myr in Waltz and from $196$ to $80$ Myr in Tango across the satellite-mass prior.

Each realization draws $r_{\rm p}$ and $\Delta v$ within their uncertainties and deprojects them with isotropic orientations. It is bound if its orbital energy is negative. We integrate each orbit with a leapfrog scheme (time step $0.05$ Myr, up to $2$ Gyr) under the gravity of the host and Chandrasekhar dynamical friction \citep{Chandra_1943},
\begin{equation}
\mathbf a_{\rm df} = -\frac{4\pi G^2 m_{\rm sat}\,\rho(r)\,\ln\Lambda}{v^3}\,\chi(X)\,\mathbf v,
\end{equation}
with $\chi(X) = {\rm erf}(X) - 2X e^{-X^2}/\sqrt{\pi}$, $X = v/[\sqrt{2}\,\sigma(r)]$, $\ln\Lambda = \ln[1 + M(<r)/m_{\rm sat}]$, and $\sigma(r)$ the velocity dispersion of the isotropic NFW halo \citep{Zentner_Bullock_2003}.
Tidal stripping drives the satellite mass toward the mass within its tidal radius, $r_{\rm t} = r\,\{m_{\rm sat}/[(2 - {\rm d}\ln M/{\rm d}\ln r)\,M(<r)]\}^{1/3}$, on the local orbital time $2\pi r/V_{\rm c}(r)$. We record a merger when the separation falls below $500$ pc, where the pair would appear as a single source at our data resolution ($0\farcs1 = 0.6$ kpc at $z = 5.5$). Orbits receding beyond $3\,r_{200}$ are, instead, counted as escaped.

Figure \ref{fig:dynamics} (panel a) answers the binding question as a function of host halo mass. 
For Tango, the requirement is very mild: binding needs $9.5 \times 10^{8} \Msun$ within $2.54$ kpc, and the pair is bound in $71-90\%$ of the realizations for every halo mass in the prior ($81\%$ marginalized). For Waltz, the requirement is $7.2 \times 10^{9} \Msun$ within $3.57$ kpc, so that the bound fraction rises from $17\%$ to $71\%$ across the prior ($48\%$ marginalized) and crosses $50\%$ near $5 \times 10^{10} \Msun$. The baryonic mass of the host matters little, because the halo sets the escape speed. The bound fraction is $0.46$ for hosts of $10^{8}-5 \times 10^{8} \Msun$ and $0.51$ for $2.5-8 \times 10^{9} \Msun$ ($0.80$ and $0.82$ for Tango). Capping the baryons at the dust-free limit of $3 \times 10^{9} \Msun$ leaves the marginalized value at $48\%$ and the crossing at $5 \times 10^{10} \Msun$. Hence, for Waltz, the binding question turns into a dynamical measurement of the halo that hosts the pair: if the pair is bound, that halo has $M_{200} \gtrsim 5 \times 10^{10} \Msun$, with $\log (M_{200}/\Msun) = 10.96^{+0.38}_{-0.50}$ across the bound realizations. This is the halo at $z = 3.9$, which is about a Gyr after any heavy seed would have formed (\S\ref{sec:summary_concl}). In the reference configuration, the companion is the satellite in every bound orbit.

Dynamical friction dominates orbital decay (panel b). The median bound orbit of Tango merges in $129$ Myr, i.e., before $z = 5.0$. For Waltz, it is $457$ Myr, i.e., by $z \approx 3.2$. Both distributions peak well before their medians, at $22$ and $231$ Myr, respectively, and have long tails. Of the bound Tango orbits, $69\%$ merge within $300$ Myr and $88\%$ within $1$ Gyr; of the Waltz ones, $24\%$ and $64\%$. 

Panels c and d show a family of elongated orbits threading the LRD, with the median one implying two passages and a merger within $129$ Myr. According to the median orbit, deprojected, the Tango companion is $2.95$ kpc from its LRD, moving at $114 \, \rm km \, s^{-1}$; the Waltz companion is $4.13$ kpc away at $265 \, \rm km \, s^{-1}$. Both median orbits are eccentric with circularities $0.68$ and $0.49$, and both pairs are within about $50$ Myr of their first pericenter, at $1.24$ kpc for Tango and $2.49$ kpc for Waltz. In summary, in the bound realizations the pair lies within about 50 Myr of pericenter, and in half of them it is approaching.

These results rest on two observables and priors: isotropic orientations, halo mass, host baryonic mass, and the choice of center. Tango's binding is robust to all of them, while Waltz's binding depends on halo mass. The merger time for any single realization is uncertain by a factor of several because the deprojected separation and tangential velocity are unknown, while the ensemble establishes the shape of the distribution.

When the cosmic dance ends, the pair has merged. A few $10^{7}-10^{8} \Msun$ of young, metal-poor stars and their gas then surround black holes that are still within a few e-foldings of their birth mass. If the companion is the halo whose radiation triggered the collapse, it ends up in the same galaxy as the black hole it helped to form, and the LRD phase ends in a merger.

\section{Summary and Conclusions}
\label{sec:summary_concl}

This study was motivated by observations  \citep{rinaldi_not_2025, Baggen_2026, Barger_2026} and a statistical analysis by \cite{Pacucci_Urry_2026}, which found that the brightest LRDs host blue companions within $2$ kpc, $2.39$ times more often than matched galaxies (significant at $3.3\sigma$), and that those companions are bluer than the companions of ordinary galaxies. That excess suggests the geometry of the ``synchronized pair'' \citep{Dijkstra_2008, Visbal_2014_pair, Regan_2017}, in which a young, star-forming halo irradiates a pristine neighbor and enables the formation of a DCBH. Photometric measurements, however, left open four questions that can only be addressed spectroscopically: whether (i) the companions are physical, (ii) their blue light is starlight or radiation reprocessed from the LRD, (iii) how massive the black holes are, and (iv) whether the pairs are gravitationally bound.

This Letter presents the only two LRD--companion pairs reported in the literature with spectra of both members: Waltz, at $z = 3.936$, observed with the NIRSpec MSA, and Tango, at $z = 5.4816$, observed with the NIRSpec IFU. Our key findings are the following.

\begin{itemize}
\item Both companions lie at the redshift of their LRD, offset by $132$ and $57 \, \rm km \, s^{-1}$ at projected separations of $3.6$ and $2.5$ kpc.

\item The two companions are physically similar, but separated by $524$ Myr of cosmic time: starbursts $\lesssim 100$ Myr old, metal-poor ($0.2 \, \rm Z_\odot$), dust-poor, dynamically cold ($\sigma_{\rm int} = 20-65 \, \rm km \, s^{-1}$), of $10^{7}-10^{8} \Msun$, and as bright as their LRDs in the rest-frame UV.

\item The LRD does not power the companion's light.

\item The LRD in Waltz hosts a black hole with a virial mass of $\approx 3 \times 10^{6} \Msun$, accreting at half the Eddington rate, only a factor of a few above the birth mass of a DCBH, inside narrow-line gas as metal-poor as the companion's and free of dust, in a host with $\Mstar < 3 \times 10^{9} \Msun$ from a dust-free, stars-only fit. If electron scattering inflates the virial masses, the black hole lies within the direct collapse birth range.

\item Tango is bound in $71-90\%$ of the orbital realizations across the halo-mass prior; Waltz is bound in more than half only if its host halo exceeds $5 \times 10^{10} \Msun$, so that its binding is a conditional constraint on that halo. Which member sits at its center is not known. Bound pairs merge within median times of $129$ and $457$ Myr (modes of $22$ and $231$ Myr).
\end{itemize}

The ages of the blue companions raise a timing question. The starbursts we observe are $\lesssim 100$ Myr old, whereas direct collapse requires gas that is (nearly) pristine (below $\sim 10^{-4}-10^{-3} \, \rm Z_\odot$, \citealt{Omukai_S_H_2008, Chon_Omukai_2020, Regan_2020_metals}). Hence, the seed must have formed before the surrounding gas was enriched: the LRDs' narrow-line regions, with [\ion{N}{2}]/H$\alpha < 0.08$ and $0.008$, are metal-poor but not pristine. Two possibilities are open. 

In the standard scenario, the seed formed at $z \gtrsim 10$, and the companion is the descendant of the first halo to collapse, forming stars in episodes of which we observe the latest.
In Waltz, knot B of the companion, with its Balmer break and $4 \times 10^{8} \Msun$ of stars, supports this interpretation: star formation beside the LRD predates the current burst. Tango may be the later stage in a second sense, since the diffuse emission around the pair and a second blue source $8.6$ kpc from the LRD (\S\ref{subsec:association}) are consistent with a system that has already absorbed an earlier companion, and its black hole is more massive and its narrow-line gas dustier ($A_V \simeq 0.9$ mag) than those of Waltz.

In the second scenario, the whole system is young: the companions assembled most of their stars within the last $\sim 100$ Myr ($\Mstar/{\rm SFR} \approx 56$ and $65$ Myr), the Waltz black hole needs $60-130$ Myr to grow from a $10^{6} \Msun$ seed to its observed mass at its observed Eddington ratio, and a hundred Myr of star formation is enough to enrich the companions to the $0.2 \, \rm Z_\odot$ we measure from their spectra, which is also the metallicity that galaxies of $10^{7}-10^{8} \Msun$ display at $z = 4-10$ \citep{Nakajima_2023, Curti_2023}, and for their supernova-driven winds to carry metals across the few kpc that separate them from their neighbors \citep{Dijkstra_2014, Agarwal_2014}.

The binding masses displayed in Fig. \ref{fig:dynamics} also add an important timing argument. Direct collapse takes place in atomic-cooling halos of $\sim 10^{7}-10^{8} \Msun$ at $z \gtrsim 10$ \citep{Bromm_Loeb_2003, Inayoshi_review_2019}, whereas if the Waltz pair is bound, it sits in a halo of $\gtrsim 5 \times 10^{10} \Msun$ at $z = 3.9$. However, these two numbers refer to epochs separated by $\sim 1$ Gyr. Early halos on rapid growth tracks of high-$\sigma$ peaks (e.g., the members of a synchronized pair) grow as $M(z) \simeq M_0 \, e^{-\alpha (z - z_0)}$ with $\alpha \approx 0.8-1$ \citep{Dekel_2013, Correa_2015}, i.e., by a factor $e^{6\alpha} \approx 100-400$ between $z_0 = 10$ and $z = 4$, while the abundance of LRDs at these redshifts independently points to hosts of this mass \citep{Pacucci_Loeb_2025}. 

If the seed formed at $z \gtrsim 10$, its halo has had a billion years to grow, and we expect a mass of $5 \times 10^{10} \Msun$ at $z = 3.9$. If instead the seed formed in the last $100-200$ Myr, its halo would have just crossed the atomic-cooling threshold, $2-4 \times 10^{8} \Msun$ at $z = 4-6$, and could not have reached $5 \times 10^{10} \Msun$ since. Hence, for Waltz, the two interpretations diverge: if the pair is bound, the seed is old; if the system is young, the pair is unbound, which its $48\%$ bound probability allows. An old seed must then have grown slowly. Reaching $3 \times 10^{6} \Msun$ takes $100-520$ Myr at the present rate for seeds of $10^{6}-10^{4} \Msun$, against the $\sim 1$ Gyr elapsed since $z \approx 10$, so the black hole was active for $10-50\%$ of its life, as the intermittent, radiation-regulated accretion of DCBHs predicts \citep{Pacucci_2015}. This argument assumes that the LRD is the central object. If the companion is central instead, the massive halo is the companion's, and a young seed in a satellite halo that it irradiates can also be bound \citep{Agarwal_2013}. Tango is less constraining: binding requires only $10^{9} \Msun$ inside 2.5 kpc (Table~\ref{tab:properties}), which an atomic-cooling halo alone could not supply but $\sim 10^{9} \Msun$ of baryons in the LRD host could, so both interpretations remain open. 

The young interpretation requires a near-pristine atomic-cooling halo beside a star-forming one at $z = 4-6$, which are expected to exist. Simulations find metal-free star formation continuing in pristine pockets at the outskirts of enriched regions well past the end of reionization \citep{Tornatore_2007, Trenti_2009, Liu_Bromm_2020, Venditti_2026}, with the sites shifting to atomic-cooling halos by $z \sim 5$ \citep{Hegde_Furlanetto_2025}. In addition, Population III candidates are now reported at $z = 6-8$ \citep{Vanzella_2023, Fujimoto_2025}. Of course, our observations cannot establish whether such a pocket existed next to the companion at the time of collapse.

Wherever the pair is bound, in most Tango realizations and in about half of the Waltz ones, the end is the same. Within a few hundred Myr, the pair merges, and the black hole gains the companion's $\sim  10^{8} \Msun$ of young, metal-poor stars. The LRD hosts are faint and hidden beneath the compact nucleus in the rest-frame optical, with only upper limits on stellar masses, so the merger adds a stellar component of $\sim 10^{8} \Msun$ to a host whose stellar mass is bounded only from above, around a black hole that is already $10^{6}-10^{7} \Msun$. Thus, the synchronized pair mechanism first generates the seed, then forms the seed's first galaxy. In this scenario, a black hole that acquires its stars after forming is overmassive by construction and descends toward the local relation as the host galaxy assembles around it. An LRD may indicate the phase between the two acts, when the black hole and its future galaxy are still apart.

If such mergers recur, the LRD phase has a natural clock. Each one adds $\sim 10^{8}\Msun$ of stars within a few hundred Myr. Over a Gyr, several would add $\sim 10^{9}\Msun$ of stars to the LRD host, along with gas to fuel in situ star formation. The accreted stars alone are not enough to carry a $10^{6}$--$10^{7}\Msun$ black hole onto the local relation, but the sequence of mergers could end the phase in which the dot outshines its galaxy, on the same Gyr timescale over which LRDs disappear below $z \approx 4$ (e.g., \citealt{Kocevski_2024, Kokorev_2024_census, Pacucci_Loeb_2025}).

Although based on only two systems, our study highlights the urgent need to investigate LRDs' immediate surroundings, which have received relatively little attention to date, with only a handful of spectroscopic observations, most of them designed for the LRD rather than its companion.

In summary, spectroscopy of LRDs' blue companions shows they are real, young, metal-poor starbursts. With Waltz and Tango, we are watching two steps of the same dance at two epochs: a heavy black hole seed beside the galaxy that may have lit its collapse, a few hundred Myr before the two become one.

\begin{acknowledgments}
F.P. acknowledges support from NASA through Chandra Award No. GO3-24087A and JWST Award No. JWST-GO-03805.010-A. P.R. acknowledges support from the University of Texas at Austin Cosmic Frontier Center. 
This work is based in part on observations made with the NASA/ESA/CSA James Webb Space Telescope. The data were obtained from the Mikulski Archive for Space Telescopes at the Space Telescope Science Institute, which is operated by the Association of Universities for Research in Astronomy, Inc., under NASA contract NAS 5-03127 for JWST. These observations are associated with programs \#1180, \#1286, and \#5664, and with the imaging programs combined in the JADES and DJA mosaics of GOODS-S (\S\ref{subsec:data}). This work also uses observations made with the NASA/ESA Hubble Space Telescope, obtained from MAST under NASA contract NAS 5-26555.
Some of the data products presented herein were retrieved from the Dawn JWST Archive (DJA), an initiative of the Cosmic Dawn Center, funded under grant DNRF140.
\end{acknowledgments}

\software{\textsc{astropy} \citep{Astropy_2013, Astropy_2018, Astropy_2022},
          \textsc{numpy} \citep{Numpy_2020},
          \textsc{scipy} \citep{Scipy_2020},
          \textsc{matplotlib} \citep{Matplotlib_2007},
          \textsc{EAZY} \citep{Brammer_2008},
          \textsc{SEP} \citep{Barbary_2016},
          \textsc{bagpipes} \citep{carnall_vandels_2019},
          \textsc{grizli} \citep{Brammer_2023}
          }

\vspace{4.5cm}
\appendix
\section{SED Fits}
\label{app:sed}

\begin{figure*}[!tp]
\centering
\includegraphics[width=\textwidth]{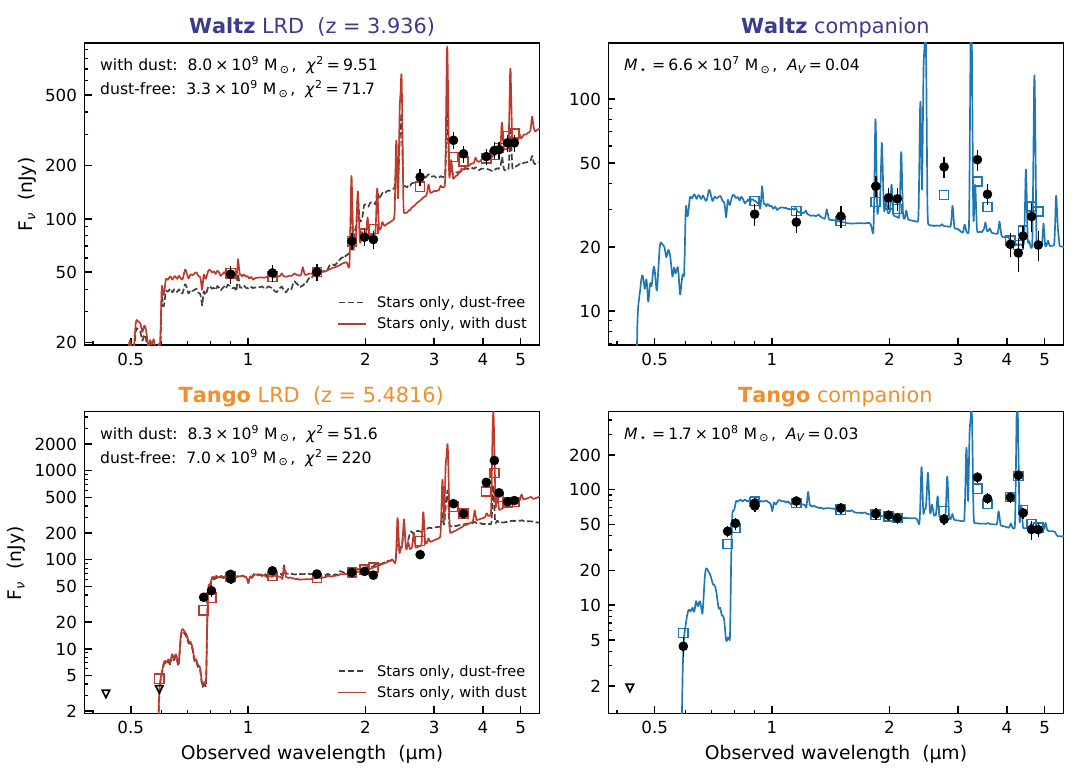}
\caption{SED fits from \textsc{EAZY} at fixed redshift (\S\ref{sec:photometry_orbits}). Filled circles are the photometry and triangles $2\sigma$ upper limits; open squares are the model fluxes, and lines the best-fit templates, smoothed to $R = 100$. Left: the LRDs, fitted with stars only, with dust (solid; $A_V = 1.4$ and $2.6$ mag) and without (dashed); labels give the mass and $\chi^2$ of each fit, and these masses are upper limits. Right: the companions (knot A for Waltz).}
\label{fig:sedfits}
\end{figure*}

Figure~\ref{fig:sedfits} shows the EAZY SED fits of \S\ref{sec:photometry_orbits}, at the redshift of each pair. Young, dust-poor templates describe the companions. In the Waltz companion, the fit falls $\approx 2\sigma$ short in the two bands boosted by [\ion{O}{3}] and H$\alpha$ (F277W and F335M), i.e., the observed lines are stronger than those of the templates. For the LRDs, stellar templates reach the red continuum only with dust, so that with the dust-free templates alone, the fits are much poorer, $\chi^2 = 72$ against $9.5$ for Waltz and $220$ against $52$ for Tango, and the masses are $3.3 \times 10^{9}$ and $7.0 \times 10^{9} \Msun$. Table~\ref{tab:properties} lists the dust-free limit for Waltz and the dust-included limit for Tango.

\bibliography{references}{}

@ARTICLE{theios_dust_2019,
       author = {{Theios}, Rachel L. and {Steidel}, Charles C. and {Strom}, Allison L. and {Rudie}, Gwen C. and {Trainor}, Ryan F. and {Reddy}, Naveen A.},
        title = "{Dust Attenuation, Star Formation, and Metallicity in z {\ensuremath{\sim}} 2-3 Galaxies from KBSS-MOSFIRE}",
      journal = {\apj},
         year = 2019,
        month = jan,
       volume = {871},
       number = {1},
          eid = {128},
        pages = {128},
          doi = {10.3847/1538-4357/aaf386},
archivePrefix = {arXiv},
       eprint = {1805.00016},
 primaryClass = {astro-ph.GA},
       adsurl = {https://ui.adsabs.harvard.edu/abs/2019ApJ...871..128T}
}

@ARTICLE{navarro-carrera_burstiness_2026,
       author = {{Navarro-Carrera}, Rafael and {Rinaldi}, Pierluigi and {Caputi}, Karina I. and {Iani}, Edoardo and {Kokorev}, Vasily and {Kerutt}, Josephine and {Cooper}, Ryan},
        title = "{Burstiness in Low Stellar Mass H{\ensuremath{\alpha}} Emitters at z {\ensuremath{\sim}} 2 and z {\ensuremath{\sim}} 4─6 from JWST Medium-band Photometry in GOODS-S}",
      journal = {\apj},
         year = 2026,
        month = jan,
       volume = {996},
       number = {1},
          eid = {70},
        pages = {70},
          doi = {10.3847/1538-4357/ae1609},
archivePrefix = {arXiv},
       eprint = {2410.23249},
 primaryClass = {astro-ph.GA},
       adsurl = {https://ui.adsabs.harvard.edu/abs/2026ApJ...996...70N}
}

@ARTICLE{KorhonenCuestas_2025,
       author = {{Korhonen Cuestas}, Nathalie A. and {Strom}, Allison L. and {Miller}, Tim B. and {Steidel}, Charles C. and {Trainor}, Ryan F. and {Rudie}, Gwen C. and {Nu{\~n}ez}, Evan Haze},
        title = "{Exploring the Relationship between Stellar Mass, Metallicity, and Star Formation Rate at z {\ensuremath{\sim}} 2.3 in KBSS-MOSFIRE}",
      journal = {\apj},
         year = 2025,
        month = may,
       volume = {984},
       number = {2},
          eid = {188},
        pages = {188},
          doi = {10.3847/1538-4357/adc5f7},
archivePrefix = {arXiv},
       eprint = {2503.10800},
 primaryClass = {astro-ph.GA},
       adsurl = {https://ui.adsabs.harvard.edu/abs/2025ApJ...984..188K}
}

@ARTICLE{rinaldi_not_2025,
       author = {{Rinaldi}, P. and {Bonaventura}, N. and {Rieke}, G.~H. and {Alberts}, S. and {Caputi}, K.~I. and {Baker}, W.~M. and {Baum}, S. and {Bhatawdekar}, R. and {Bunker}, A.~J. and {Carniani}, S. and {Curtis-Lake}, E. and {D'Eugenio}, F. and {Egami}, E. and {Ji}, Z. and {Johnson}, B.~D. and {Hainline}, K. and {Helton}, J.~M. and {Lin}, X. and {Lyu}, J. and {Ma}, Z. and {Maiolino}, R. and {P{\'e}rez-Gonz{\'a}lez}, P.~G. and {Rieke}, M. and {Robertson}, B.~E. and {Shivaei}, I. and {Stone}, M. and {Sun}, Y. and {Tacchella}, S. and {{\"U}bler}, H. and {Williams}, C.~C. and {Willmer}, C.~N.~A. and {Willott}, C. and {Zhang}, J. and {Zhu}, Y.},
        title = "{Not Just a Dot: The Complex UV Morphology and Underlying Properties of Little Red Dots}",
      journal = {\apj},
         year = 2025,
        month = oct,
       volume = {992},
       number = {1},
          eid = {71},
        pages = {71},
          doi = {10.3847/1538-4357/adfa10},
archivePrefix = {arXiv},
       eprint = {2411.14383},
 primaryClass = {astro-ph.GA},
       adsurl = {https://ui.adsabs.harvard.edu/abs/2025ApJ...992...71R}
}

@ARTICLE{juodzbalis_incomplete_2026,
       author = {{Juod{\v{z}}balis}, Ignas and {Ji}, Xihan and {D'Eugenio}, Francesco and {Scholtz}, Jan and {Maiolino}, Roberto and {Stoffers}, Amanda and {Marconi}, Alessandro and {Bertola}, Elena and {Bunker}, Andrew J. and {Carniani}, Stefano and et al.},
        title = "{An (in)complete NIRSpec census of Balmer absorption in Type 1 AGN - radiation-driven outflows in little red dots, quasars and variable stars}",
      journal = {\mnras},
         year = 2026,
        month = nov,
          doi = {10.1093/mnras/stag2131},
archivePrefix = {arXiv},
       eprint = {2607.19057},
 primaryClass = {astro-ph.GA},
       adsurl = {https://scixplorer.org/abs/2026MNRAS.tmp.1966J}
}

@ARTICLE{leja_how_2019,
       author = {{Leja}, Joel and {Carnall}, Adam C. and {Johnson}, Benjamin D. and {Conroy}, Charlie and {Speagle}, Joshua S.},
        title = "{How to Measure Galaxy Star Formation Histories. II. Nonparametric Models}",
      journal = {\apj},
         year = 2019,
        month = may,
       volume = {876},
       number = {1},
          eid = {3},
        pages = {3},
          doi = {10.3847/1538-4357/ab133c},
archivePrefix = {arXiv},
       eprint = {1811.03637},
 primaryClass = {astro-ph.GA},
       adsurl = {https://ui.adsabs.harvard.edu/abs/2019ApJ...876....3L}
}

@ARTICLE{carnall_vandels_2019,
       author = {{Carnall}, A.~C. and {McLure}, R.~J. and {Dunlop}, J.~S. and {Cullen}, F. and {McLeod}, D.~J. and {Wild}, V. and {Johnson}, B.~D. and {Appleby}, S. and {Dav{\'e}}, R. and {Amorin}, R. and {Bolzonella}, M. and {Castellano}, M. and {Cimatti}, A. and {Cucciati}, O. and {Gargiulo}, A. and {Garilli}, B. and {Marchi}, F. and {Pentericci}, L. and {Pozzetti}, L. and {Schreiber}, C. and {Talia}, M. and {Zamorani}, G.},
        title = "{The VANDELS survey: the star-formation histories of massive quiescent galaxies at 1.0 < z < 1.3}",
      journal = {\mnras},
         year = 2019,
        month = nov,
       volume = {490},
       number = {1},
        pages = {417-439},
          doi = {10.1093/mnras/stz2544},
archivePrefix = {arXiv},
       eprint = {1903.11082},
 primaryClass = {astro-ph.GA},
       adsurl = {https://ui.adsabs.harvard.edu/abs/2019MNRAS.490..417C}
}

@ARTICLE{rinaldi_way_2026,
       author = {{Rinaldi}, Pierluigi and {Hainline}, Kevin and {D'Eugenio}, Francesco and {P{\'e}rez-Gonz{\'a}lez}, Pablo G. and {Eisenstein}, Daniel J. and {Willmer}, Christopher N.~A. and {Carreira}, Courtney and {Robertson}, Brant and {Johnson}, Benjamin D. and {Alberts}, Stacey and {Baker}, William M. and {Bunker}, Andrew J. and {Carniani}, Stefano and {Egami}, Eiichi and {Helton}, Jakob M. and {Ji}, Zhiyuan and {Juod{\v{z}}balis}, Ignas and {Lin}, Xiaojing and {Lyu}, Jianwei and {Ma}, Zheng and {Maiolino}, Roberto and {Parlanti}, Eleonora and {Scholtz}, Jan and {Sun}, Yang and {Tacchella}, Sandro and {Venturi}, Giacomo and {Williams}, Christina C. and {Willott}, Chris and {Witstok}, Joris and {Wu}, Zihao},
        title = "{The Way We Tally Becomes the Tale: the Impact of Selection Strategies on the Inferred Evolution of Little Red Dots Across Cosmic Time}",
      journal = {arXiv e-prints},
         year = 2026,
        month = apr,
          eid = {arXiv:2604.07138},
        pages = {arXiv:2604.07138},
          doi = {10.48550/arXiv.2604.07138},
archivePrefix = {arXiv},
       eprint = {2604.07138},
 primaryClass = {astro-ph.GA},
       adsurl = {https://ui.adsabs.harvard.edu/abs/2026arXiv260407138R}
}

@ARTICLE{Chon_Omukai_2020,
       author = {{Chon}, Sunmyon and {Omukai}, Kazuyuki},
        title = "{Supermassive star formation via super competitive accretion in slightly metal-enriched clouds}",
      journal = {\mnras},
         year = 2020,
        month = may,
       volume = {494},
       number = {2},
        pages = {2851-2860},
          doi = {10.1093/mnras/staa863},
archivePrefix = {arXiv},
       eprint = {2001.06491},
 primaryClass = {astro-ph.GA},
       adsurl = {https://ui.adsabs.harvard.edu/abs/2020MNRAS.494.2851C}
}

@ARTICLE{Omukai_S_H_2008,
       author = {{Omukai}, K. and {Schneider}, R. and {Haiman}, Z.},
        title = "{Can Supermassive Black Holes Form in Metal-enriched High-Redshift Protogalaxies?}",
      journal = {\apj},
         year = 2008,
        month = oct,
       volume = {686},
       number = {2},
        pages = {801-814},
          doi = {10.1086/591636},
archivePrefix = {arXiv},
       eprint = {0804.3141},
 primaryClass = {astro-ph},
       adsurl = {https://ui.adsabs.harvard.edu/abs/2008ApJ...686..801O}
}

@ARTICLE{Brammer_2008,
       author = {{Brammer}, Gabriel B. and {van Dokkum}, Pieter G. and {Coppi}, Paolo},
        title = "{EAZY: A Fast, Public Photometric Redshift Code}",
      journal = {\apj},
         year = 2008,
        month = oct,
       volume = {686},
       number = {2},
        pages = {1503-1513},
          doi = {10.1086/591786},
archivePrefix = {arXiv},
       eprint = {0807.1533},
 primaryClass = {astro-ph},
       adsurl = {https://ui.adsabs.harvard.edu/abs/2008ApJ...686.1503B}
}

@ARTICLE{Decarli_2024,
       author = {{Decarli}, Roberto and {Loiacono}, Federica and {Farina}, Emanuele Paolo and {Dotti}, Massimo and {Lupi}, Alessandro and {Meyer}, Romain A. and {Mignoli}, Marco and {Pensabene}, Antonio and {Strauss}, Michael A. and {Venemans}, Bram and {Yang}, Jinyi and {Walter}, Fabian and {Wolf}, Julien and {Ba{\~n}ados}, Eduardo and {Blecha}, Laura and {Bosman}, Sarah and {Carilli}, Chris L. and {Comastri}, Andrea and {Connor}, Thomas and {Costa}, Tiago and {Eilers}, Anna-Christina and {Fan}, Xiaohui and {Gilli}, Roberto and {Jun}, Hyunsung D. and {Liu}, Weizhe and {Marshall}, Madeline A. and {Mazzucchelli}, Chiara and {Neeleman}, Marcel and {Onoue}, Masafusa and {Overzier}, Roderik and {Pudoka}, Maria Anne and {Riechers}, Dominik A. and {Rix}, Hans-Walter and {Schindler}, Jan-Torge and {Trakhtenbrot}, Benny and {Trebitsch}, Maxime and {Vestergaard}, Marianne and {Volonteri}, Marta and {Wang}, Feige and {Zhang}, Huanian and {Zou}, Siwei},
        title = "{A quasar-galaxy merger at z {\ensuremath{\sim}} 6.2: Rapid host growth via the accretion of two massive satellite galaxies}",
      journal = {\aap},
         year = 2024,
        month = sep,
       volume = {689},
          eid = {A219},
        pages = {A219},
          doi = {10.1051/0004-6361/202449239},
archivePrefix = {arXiv},
       eprint = {2406.06697},
 primaryClass = {astro-ph.GA},
       adsurl = {https://ui.adsabs.harvard.edu/abs/2024A&A...689A.219D}
}

@ARTICLE{Marshall_2023,
       author = {{Marshall}, Madeline A. and {Perna}, Michele and {Willott}, Chris J. and {Maiolino}, Roberto and {Scholtz}, Jan and {{\"U}bler}, Hannah and {Carniani}, Stefano and {Arribas}, Santiago and {L{\"u}tzgendorf}, Nora and {Bunker}, Andrew J. and {Charlot}, Stephane and {Ferruit}, Pierre and {Jakobsen}, Peter and {Rix}, Hans-Walter and {Rodr{\'\i}guez Del Pino}, Bruno and {B{\"o}ker}, Torsten and {Cameron}, Alex J. and {Cresci}, Giovanni and {Curtis-Lake}, Emma and {Jones}, Gareth C. and {Kumari}, Nimisha and {P{\'e}rez-Gonz{\'a}lez}, Pablo G. and {Reed}, Sophie L.},
        title = "{GA-NIFS: Black hole and host galaxy properties of two z ≃ 6.8 quasars from the NIRSpec IFU}",
      journal = {\aap},
         year = 2023,
        month = oct,
       volume = {678},
          eid = {A191},
        pages = {A191},
          doi = {10.1051/0004-6361/202346113},
archivePrefix = {arXiv},
       eprint = {2302.04795},
 primaryClass = {astro-ph.GA},
       adsurl = {https://ui.adsabs.harvard.edu/abs/2023A&A...678A.191M}
}

@ARTICLE{Husemann_2013,
       author = {{Husemann}, B. and {Wisotzki}, L. and {S{\'a}nchez}, S.~F. and {Jahnke}, K.},
        title = "{The properties of the extended warm ionised gas around low-redshift QSOs and the lack of extended high-velocity outflows}",
      journal = {\aap},
         year = 2013,
        month = jan,
       volume = {549},
          eid = {A43},
        pages = {A43},
          doi = {10.1051/0004-6361/201220076},
archivePrefix = {arXiv},
       eprint = {1210.0566},
 primaryClass = {astro-ph.CO},
       adsurl = {https://ui.adsabs.harvard.edu/abs/2013A&A...549A..43H}
}

@ARTICLE{Fujimoto_2025,
       author = {{Fujimoto}, Seiji and {Naidu}, Rohan P. and {Chisholm}, John and {Atek}, Hakim and {Endsley}, Ryan and {Kokorev}, Vasily and {Furtak}, Lukas J. and {Pan}, Richard and {Liu}, Boyuan and {Bromm}, Volker and {Venditti}, Alessandra and {Visbal}, Eli and {Sarmento}, Richard and {Weibel}, Andrea and {Oesch}, Pascal A. and {Brammer}, Gabriel and {Schaerer}, Daniel and {Adamo}, Angela and {Berg}, Danielle A. and {Bezanson}, Rachel and {Bouwens}, Rychard and {Chemerynska}, Iryna and {Claeyssens}, Ad{\'e}la{\"\i}de and {Dessauges-Zavadsky}, Miroslava and {Frebel}, Anna and {Korber}, Damien and {Labbe}, Ivo and {Marques-Chaves}, Rui and {Matthee}, Jorryt and {McQuinn}, Kristen B.~W. and {Mu{\~n}oz}, Julian B. and {Natarajan}, Priyamvada and {Saldana-Lopez}, Alberto and {Suess}, Katherine A. and {Volonteri}, Marta and {Zitrin}, Adi},
        title = "{GLIMPSE: An Ultrafaint ≃{}10$^{5}$ M$_{{\ensuremath{\odot}}}$ Pop III Galaxy Candidate and First Constraints on the Pop III UV Luminosity Function at z ≃ 6─7}",
      journal = {\apj},
         year = 2025,
        month = aug,
       volume = {989},
       number = {1},
          eid = {46},
        pages = {46},
          doi = {10.3847/1538-4357/ade9a1},
archivePrefix = {arXiv},
       eprint = {2501.11678},
 primaryClass = {astro-ph.GA},
       adsurl = {https://ui.adsabs.harvard.edu/abs/2025ApJ...989...46F}
}

@ARTICLE{Vanzella_2023,
       author = {{Vanzella}, E. and {Loiacono}, F. and {Bergamini}, P. and {Me{\v{s}}tri{\'c}}, U. and {Castellano}, M. and {Rosati}, P. and {Meneghetti}, M. and {Grillo}, C. and {Calura}, F. and {Mignoli}, M. and {Brada{\v{c}}}, M. and {Adamo}, A. and {Rihtar{\v{s}}i{\v{c}}}, G. and {Dickinson}, M. and {Gronke}, M. and {Zanella}, A. and {Annibali}, F. and {Willott}, C. and {Messa}, M. and {Sani}, E. and {Acebron}, A. and {Bolamperti}, A. and {Comastri}, A. and {Gilli}, R. and {Caputi}, K.~I. and {Ricotti}, M. and {Gruppioni}, C. and {Ravindranath}, S. and {Mercurio}, A. and {Strait}, V. and {Martis}, N. and {Pascale}, R. and {Caminha}, G.~B. and {Annunziatella}, M. and {Nonino}, M.},
        title = "{An extremely metal-poor star complex in the reionization era: Approaching Population III stars with JWST}",
      journal = {\aap},
         year = 2023,
        month = oct,
       volume = {678},
          eid = {A173},
        pages = {A173},
          doi = {10.1051/0004-6361/202346981},
archivePrefix = {arXiv},
       eprint = {2305.14413},
 primaryClass = {astro-ph.GA},
       adsurl = {https://ui.adsabs.harvard.edu/abs/2023A&A...678A.173V}
}

@ARTICLE{Hegde_Furlanetto_2025,
       author = {{Hegde}, Sahil and {Furlanetto}, Steven R.},
        title = "{Efficient semi-analytic modelling of Pop III star formation from Cosmic Dawn to Reionization}",
      journal = {The Open Journal of Astrophysics},
         year = 2025,
        month = oct,
       volume = {8},
          eid = {147},
        pages = {147},
          doi = {10.33232/001c.145070},
archivePrefix = {arXiv},
       eprint = {2507.19581},
 primaryClass = {astro-ph.GA},
       adsurl = {https://ui.adsabs.harvard.edu/abs/2025OJAp....8E.147H}
}

@ARTICLE{Venditti_2026,
       author = {{Venditti}, Alessandra and {Schaerer}, Daniel and {Zackrisson}, Erik and {Asada}, Yoshihisa and {Katz}, Harley and {Salvadori}, Stefania and {Vanzella}, Eros and {Mu{\~n}oz}, Julian B. and {Storck}, Anatole and {Bunker}, Andrew J. and {Trinca}, Alessandro and {Scholte}, Dirk and {Pacucci}, Fabio and {P{\'e}rez-Gonz{\'a}lez}, Pablo G. and {Fujimoto}, Seiji and {Charbonnel}, Corinne and {Maiolino}, Roberto and {Ferrara}, Andrea and {Giavalisco}, Mauro and {Schneider}, Raffaella and {Baggen}, Josephine and {Atek}, Hakim and {Bromm}, Volker and {Caputi}, Karina and {Ciesla}, Laure and {Dayal}, Pratika and {Kobayashi}, Chiaki and {Castellano}, Marco and {Santini}, Paola},
        title = "{How can we finally see the first light? Status and perspective in the search for Population III stars}",
      journal = {The Open Journal of Astrophysics},
         year = 2026,
        month = aug,
       volume = {9},
        pages = {67811},
          doi = {10.33232/001c.167811},
archivePrefix = {arXiv},
       eprint = {2607.00167},
 primaryClass = {astro-ph.GA},
       adsurl = {https://ui.adsabs.harvard.edu/abs/2026OJAp....967811V}
}

@ARTICLE{Liu_Bromm_2020,
       author = {{Liu}, Boyuan and {Bromm}, Volker},
        title = "{When did Population III star formation end?}",
      journal = {\mnras},
         year = 2020,
        month = sep,
       volume = {497},
       number = {3},
        pages = {2839-2854},
          doi = {10.1093/mnras/staa2143},
archivePrefix = {arXiv},
       eprint = {2006.15260},
 primaryClass = {astro-ph.GA},
       adsurl = {https://ui.adsabs.harvard.edu/abs/2020MNRAS.497.2839L}
}

@ARTICLE{Trenti_2009,
       author = {{Trenti}, Michele and {Stiavelli}, Massimo and {Shull}, J. Michael},
        title = "{Metal-free Gas Supply at the Edge of Reionization: Late-epoch Population III Star Formation}",
      journal = {\apj},
         year = 2009,
        month = aug,
       volume = {700},
       number = {2},
        pages = {1672-1679},
          doi = {10.1088/0004-637X/700/2/1672},
archivePrefix = {arXiv},
       eprint = {0905.4504},
 primaryClass = {astro-ph.CO},
       adsurl = {https://ui.adsabs.harvard.edu/abs/2009ApJ...700.1672T}
}

@ARTICLE{Tornatore_2007,
       author = {{Tornatore}, Luca and {Ferrara}, Andrea and {Schneider}, Raffaella},
        title = "{Population III stars: hidden or disappeared?}",
      journal = {\mnras},
         year = 2007,
        month = dec,
       volume = {382},
       number = {3},
        pages = {945-950},
          doi = {10.1111/j.1365-2966.2007.12215.x},
archivePrefix = {arXiv},
       eprint = {0707.1433},
 primaryClass = {astro-ph},
       adsurl = {https://ui.adsabs.harvard.edu/abs/2007MNRAS.382..945T}
}

@ARTICLE{Regan_2020_metals,
       author = {{Regan}, John A. and {Haiman}, Zolt{\'a}n and {Wise}, John H. and {O'Shea}, Brian W. and {Norman}, Michael L.},
        title = "{Massive Star Formation in Metal-Enriched Haloes at High Redshift}",
      journal = {The Open Journal of Astrophysics},
         year = 2020,
        month = aug,
       volume = {3},
        pages = {E9},
          doi = {10.21105/astro.2006.14625},
archivePrefix = {arXiv},
       eprint = {2006.14625},
 primaryClass = {astro-ph.GA},
       adsurl = {https://ui.adsabs.harvard.edu/abs/2020OJAp....3E...9R}
}

@ARTICLE{Curti_2020,
       author = {{Curti}, Mirko and {Mannucci}, Filippo and {Cresci}, Giovanni and {Maiolino}, Roberto},
        title = "{The mass-metallicity and the fundamental metallicity relation revisited on a fully T$_{e}$-based abundance scale for galaxies}",
      journal = {\mnras},
         year = 2020,
        month = jan,
       volume = {491},
       number = {1},
        pages = {944-964},
          doi = {10.1093/mnras/stz2910},
archivePrefix = {arXiv},
       eprint = {1910.00597},
 primaryClass = {astro-ph.GA},
       adsurl = {https://ui.adsabs.harvard.edu/abs/2020MNRAS.491..944C}
}

@ARTICLE{Eldridge_2017,
       author = {{Eldridge}, J.~J. and {Stanway}, E.~R. and {Xiao}, L. and {McClelland}, L.~A.~S. and {Taylor}, G. and {Ng}, M. and {Greis}, S.~M.~L. and {Bray}, J.~C.},
        title = "{Binary Population and Spectral Synthesis Version 2.1: Construction, Observational Verification, and New Results}",
      journal = {\pasa},
         year = 2017,
        month = nov,
       volume = {34},
          eid = {e058},
        pages = {e058},
          doi = {10.1017/pasa.2017.51},
archivePrefix = {arXiv},
       eprint = {1710.02154},
 primaryClass = {astro-ph.SR},
       adsurl = {https://ui.adsabs.harvard.edu/abs/2017PASA...34...58E}
}

@ARTICLE{Leitherer_1999,
       author = {{Leitherer}, Claus and {Schaerer}, Daniel and {Goldader}, Jeffrey D. and {Delgado}, Rosa M. Gonz{\'a}lez and {Robert}, Carmelle and {Kune}, Denis Foo and {de Mello}, Du{\'\i}lia F. and {Devost}, Daniel and {Heckman}, Timothy M.},
        title = "{Starburst99: Synthesis Models for Galaxies with Active Star Formation}",
      journal = {\apjs},
         year = 1999,
        month = jul,
       volume = {123},
       number = {1},
        pages = {3-40},
          doi = {10.1086/313233},
archivePrefix = {arXiv},
       eprint = {astro-ph/9902334},
 primaryClass = {astro-ph},
       adsurl = {https://ui.adsabs.harvard.edu/abs/1999ApJS..123....3L}
}

@BOOK{Osterbrock_Ferland_2006,
       author = {{Osterbrock}, Donald E. and {Ferland}, Gary J.},
        title = "{Astrophysics of gaseous nebulae and active galactic nuclei}",
         year = 2006,
       adsurl = {https://ui.adsabs.harvard.edu/abs/2006agna.book.....O},
       publisher = "{University Science Books}"
}

@ARTICLE{Ventou_2017,
       author = {{Ventou}, E. and {Contini}, T. and {Bouch{\'e}}, N. and {Epinat}, B. and {Brinchmann}, J. and {Bacon}, R. and {Inami}, H. and {Lam}, D. and {Drake}, A. and {Garel}, T. and {Michel-Dansac}, L. and {Pello}, R. and {Steinmetz}, M. and {Weilbacher}, P.~M. and {Wisotzki}, L. and {Carollo}, M.},
        title = "{The MUSE Hubble Ultra Deep Field Survey. IX. Evolution of galaxy merger fraction since z {\ensuremath{\approx}} 6}",
      journal = {\aap},
         year = 2017,
        month = dec,
       volume = {608},
          eid = {A9},
        pages = {A9},
          doi = {10.1051/0004-6361/201731586},
archivePrefix = {arXiv},
       eprint = {1711.00423},
 primaryClass = {astro-ph.GA},
       adsurl = {https://ui.adsabs.harvard.edu/abs/2017A&A...608A...9V}
}

@ARTICLE{Helton_2024,
       author = {{Helton}, Jakob M. and {Sun}, Fengwu and {Woodrum}, Charity and {Hainline}, Kevin N. and {Willmer}, Christopher N.~A. and {Rieke}, Marcia J. and {Rieke}, George H. and {Alberts}, Stacey and {Eisenstein}, Daniel J. and {Tacchella}, Sandro and {Robertson}, Brant and {Johnson}, Benjamin D. and {Baker}, William M. and {Bhatawdekar}, Rachana and {Bunker}, Andrew J. and {Chen}, Zuyi and {Egami}, Eiichi and {Ji}, Zhiyuan and {Maiolino}, Roberto and {Willott}, Chris and {Witstok}, Joris},
        title = "{Identification of High-redshift Galaxy Overdensities in GOODS-N and GOODS-S}",
      journal = {\apj},
         year = 2024,
        month = oct,
       volume = {974},
       number = {1},
          eid = {41},
        pages = {41},
          doi = {10.3847/1538-4357/ad6867},
archivePrefix = {arXiv},
       eprint = {2311.04270},
 primaryClass = {astro-ph.GA},
       adsurl = {https://ui.adsabs.harvard.edu/abs/2024ApJ...974...41H}
}

@ARTICLE{Ishikawa_2026,
       author = {{Ishikawa}, Yuzo and {Eilers}, Anna-Christina and {Naidu}, Rohan P. and {Matthee}, Jorryt and {Bordoloi}, Rongmon and {Chisholm}, John and {Greene}, Jenny E. and {Ma}, Yilun and {Oesch}, Pascal A. and {Sun}, Wendy Q. and {Torralba}, Alberto and {Weaver}, John R. and {Wuyts}, Stijn and {Xiao}, Mengyuan},
        title = "{Spatial decomposition of Little Red Dots with JWST/NIRSpec IFU into broad-line red cores and narrow-line blue host galaxies}",
      journal = {arXiv e-prints},
         year = 2026,
        month = jul,
          eid = {arXiv:2607.09647},
        pages = {arXiv:2607.09647},
          doi = {10.48550/arXiv.2607.09647},
archivePrefix = {arXiv},
       eprint = {2607.09647},
 primaryClass = {astro-ph.GA},
       adsurl = {https://ui.adsabs.harvard.edu/abs/2026arXiv260709647I}
}

@ARTICLE{Matthee_2026,
       author = {{Matthee}, Jorryt and {Torralba}, Alberto and {Pezzulli}, Gabriele and {Naidu}, Rohan P. and {Chisholm}, John and {Mascia}, Sara and {Greene}, Jenny E. and {Ishikawa}, Yuzo and {Gronke}, Max and {Wuyts}, Stijn and {Bordoloi}, Rongmon and {Brammer}, Gabriel and {Chang}, Seok-Jun and {Eilers}, Anna-Christina and {de Graaff}, Anna and {Hviding}, Raphael E. and {Iani}, Edoardo and {Illingworth}, Garth and {Kashino}, Daichi and {Labbe}, Ivo and {Ma}, Yilun and {Maseda}, Michael V. and {Meyer}, Romain and {Nelson}, Erica and {Oesch}, Pascal and {Xiao}, Mengyuan},
        title = "{The Engine and its Flows: Little Red Dot spectra are shaped by the column densities of their gas envelopes}",
      journal = {arXiv e-prints},
         year = 2026,
        month = mar,
          eid = {arXiv:2603.17667},
        pages = {arXiv:2603.17667},
          doi = {10.48550/arXiv.2603.17667},
archivePrefix = {arXiv},
       eprint = {2603.17667},
 primaryClass = {astro-ph.GA},
       adsurl = {https://ui.adsabs.harvard.edu/abs/2026arXiv260317667M}
}

@ARTICLE{Valentino_2023,
       author = {{Valentino}, Francesco and {Brammer}, Gabriel and {Gould}, Katriona M.~L. and {Kokorev}, Vasily and {Fujimoto}, Seiji and {Jespersen}, Christian Kragh and {Vijayan}, Aswin P. and {Weaver}, John R. and {Ito}, Kei and {Tanaka}, Masayuki and {Ilbert}, Olivier and {Magdis}, Georgios E. and {Whitaker}, Katherine E. and {Faisst}, Andreas L. and {Gallazzi}, Anna and {Gillman}, Steven and {Gim{\'e}nez-Arteaga}, Clara and {G{\'o}mez-Guijarro}, Carlos and {Kubo}, Mariko and {Heintz}, Kasper E. and {Hirschmann}, Michaela and {Oesch}, Pascal and {Onodera}, Masato and {Rizzo}, Francesca and {Lee}, Minju and {Strait}, Victoria and {Toft}, Sune},
        title = "{An Atlas of Color-selected Quiescent Galaxies at z > 3 in Public JWST Fields}",
      journal = {\apj},
         year = 2023,
        month = apr,
       volume = {947},
       number = {1},
          eid = {20},
        pages = {20},
          doi = {10.3847/1538-4357/acbefa},
archivePrefix = {arXiv},
       eprint = {2302.10936},
 primaryClass = {astro-ph.GA},
       adsurl = {https://ui.adsabs.harvard.edu/abs/2023ApJ...947...20V}
}

@ARTICLE{JADES_2025,
       author = {{D'Eugenio}, Francesco and {Cameron}, Alex J. and {Scholtz}, Jan and {Carniani}, Stefano and {Willott}, Chris J. and {Curtis-Lake}, Emma and {Bunker}, Andrew J. and {Parlanti}, Eleonora and {Maiolino}, Roberto and {Willmer}, Christopher N.~A. and {Jakobsen}, Peter and {Robertson}, Brant E. and {Johnson}, Benjamin D. and {Tacchella}, Sandro and {Cargile}, Phillip A. and {Rawle}, Tim and {Arribas}, Santiago and {Chevallard}, Jacopo and {Curti}, Mirko and {Egami}, Eiichi and {Eisenstein}, Daniel J. and {Kumari}, Nimisha and {Looser}, Tobias J. and {Rieke}, Marcia J. and {Rodr{\'\i}guez Del Pino}, Bruno and {Saxena}, Aayush and {{\"U}bler}, Hannah and {Venturi}, Giacomo and {Witstok}, Joris and {Baker}, William M. and {Bhatawdekar}, Rachana and {Bonaventura}, Nina and {Boyett}, Kristan and {Charlot}, Stephane and {Danhaive}, A. Lola and {Hainline}, Kevin N. and {Hausen}, Ryan and {Helton}, Jakob M. and {Ji}, Xihan and {Ji}, Zhiyuan and {Jones}, Gareth C. and {Juod{\v{z}}balis}, Ignas and {Maseda}, Michael V. and {P{\'e}rez-Gonz{\'a}lez}, Pablo G. and {Perna}, Michele and {Pusk{\'a}s}, D{\'a}vid and {Shivaei}, Irene and {Silcock}, Maddie S. and {Simmonds}, Charlotte and {Smit}, Renske and {Sun}, Fengwu and {Villanueva}, Natalia C. and {Williams}, Christina C. and {Zhu}, Yongda},
        title = "{JADES Data Release 3: NIRSpec/Microshutter Assembly Spectroscopy for 4000 Galaxies in the GOODS Fields}",
      journal = {\apjs},
         year = 2025,
        month = mar,
       volume = {277},
       number = {1},
          eid = {4},
        pages = {4},
          doi = {10.3847/1538-4365/ada148},
archivePrefix = {arXiv},
       eprint = {2404.06531},
 primaryClass = {astro-ph.GA},
       adsurl = {https://ui.adsabs.harvard.edu/abs/2025ApJS..277....4D}
}

@ARTICLE{Correa_2015,
       author = {{Correa}, Camila A. and {Wyithe}, J. Stuart B. and {Schaye}, Joop and {Duffy}, Alan R.},
        title = "{The accretion history of dark matter haloes - I. The physical origin of the universal function}",
      journal = {\mnras},
         year = 2015,
        month = jun,
       volume = {450},
       number = {2},
        pages = {1514-1520},
          doi = {10.1093/mnras/stv689},
archivePrefix = {arXiv},
       eprint = {1409.5228},
 primaryClass = {astro-ph.GA},
       adsurl = {https://ui.adsabs.harvard.edu/abs/2015MNRAS.450.1514C}
}

@ARTICLE{Dekel_2013,
       author = {{Dekel}, A. and {Zolotov}, A. and {Tweed}, D. and {Cacciato}, M. and {Ceverino}, D. and {Primack}, J.~R.},
        title = "{Toy models for galaxy formation versus simulations}",
      journal = {\mnras},
         year = 2013,
        month = oct,
       volume = {435},
       number = {2},
        pages = {999-1019},
          doi = {10.1093/mnras/stt1338},
archivePrefix = {arXiv},
       eprint = {1303.3009},
 primaryClass = {astro-ph.CO},
       adsurl = {https://ui.adsabs.harvard.edu/abs/2013MNRAS.435..999D}
}

@ARTICLE{Pacucci_Urry_2026,
       author = {{Pacucci}, Fabio and {Urry}, C. Megan},
        title = "{More Numerous and Bluer: Companions of the Brightest Little Red Dots Differ from Control Galaxies at > 3{\ensuremath{\sigma}}, Hinting at Synchronized Black Hole Seed Formation}",
      journal = {\apjl},
         year = 2026,
        month = oct,
       volume = {1010},
       number = {1},
          eid = {L2},
        pages = {L2},
          doi = {10.3847/2041-8213/aeaa52},
       adsurl = {https://ui.adsabs.harvard.edu/abs/2026ApJ..1010L...2P}
}

@ARTICLE{Golubchik_2026,
       author = {{Golubchik}, Miriam and {Furtak}, Lukas J. and {Allingham}, Joseph F.~V. and {Zitrin}, Adi and {Akins}, Hollis B. and {Kokorev}, Vasily and {Fujimoto}, Seiji and {Abdurro'uf} and {Amor{\'\i}n}, Ricardo O. and {Bauer}, Franz E. and {Bezanson}, Rachel and {Brada{\v{c}}}, Marusa and {Bradley}, Larry D. and {Brammer}, Gabriel B. and {Chisholm}, John and {Coe}, Dan and {Conselice}, Christopher J. and {Dayal}, Pratika and {Dessauges-Zavadsky}, Miroslava and {Diego}, Jose M. and {Faisst}, Andreas L. and {Fei}, Qinyue and {Ferguson}, Henry C. and {Finkelstein}, Steven L. and {Frye}, Brenda L. and {Gonz{\'a}lez-Otero}, Mauro and {Greene}, Jenny E. and {Harikane}, Yuichi and {Hsiao}, Tiger Yu-Yang and {Inayoshi}, Kohei and {Jim{\'e}nez-Teja}, Yolanda and {Knudsen}, Kirsten and {Koekemoer}, Anton M. and {Labb{\'e}}, Ivo and {Lucas}, Ray A. and {Magdis}, Georgios E. and {Matthee}, Jorryt and {Messa}, Matteo and {Naidu}, Rohan P. and {Nakane}, Minami and {Noirot}, Ga{\"e}l and {Pan}, Richard and {Papovich}, Casey and {Richard}, Johan and {Ricotti}, Massimo and {Robbins}, Luke and {Stark}, Daniel P. and {Sun}, Fengwu and {Treu}, Tommaso and {Tripodi}, Roberta and {Vanzella}, Eros and {Willott}, Chris and {Windhorst}, Rogier A.},
        title = "{VENUS: When red meets blue: A multiply imaged little red dot with an apparent blue companion behind the galaxy cluster Abell 383}",
      journal = {\aap},
         year = 2026,
        month = jun,
       volume = {710},
          eid = {A226},
        pages = {A226},
          doi = {10.1051/0004-6361/202558407},
archivePrefix = {arXiv},
       eprint = {2512.02117},
 primaryClass = {astro-ph.GA},
       adsurl = {https://ui.adsabs.harvard.edu/abs/2026A&A...710A.226G}
}

@ARTICLE{Rieke_2023_JADES,
       author = {{Rieke}, Marcia J. and {Robertson}, Brant and {Tacchella}, Sandro and {Hainline}, Kevin and {Johnson}, Benjamin D. and {Hausen}, Ryan and {Ji}, Zhiyuan and {Willmer}, Christopher N.~A. and {Eisenstein}, Daniel J. and {Pusk{\'a}s}, D{\'a}vid and {Alberts}, Stacey and {Arribas}, Santiago and {Baker}, William M. and {Baum}, Stefi and {Bhatawdekar}, Rachana and {Bonaventura}, Nina and {Boyett}, Kristan and {Bunker}, Andrew J. and {Cameron}, Alex J. and {Carniani}, Stefano and {Charlot}, Stephane and {Chevallard}, Jacopo and {Chen}, Zuyi and {Curti}, Mirko and {Curtis-Lake}, Emma and {Danhaive}, A. Lola and {DeCoursey}, Christa and {Dressler}, Alan and {Egami}, Eiichi and {Endsley}, Ryan and {Helton}, Jakob M. and {Hviding}, Raphael E. and {Kumari}, Nimisha and {Looser}, Tobias J. and {Lyu}, Jianwei and {Maiolino}, Roberto and {Maseda}, Michael V. and {Nelson}, Erica J. and {Rieke}, George and {Rix}, Hans-Walter and {Sandles}, Lester and {Saxena}, Aayush and {Sharpe}, Katherine and {Shivaei}, Irene and {Skarbinski}, Maya and {Smit}, Renske and {Stark}, Daniel P. and {Stone}, Meredith and {Suess}, Katherine A. and {Sun}, Fengwu and {Topping}, Michael and {{\"U}bler}, Hannah and {Villanueva}, Natalia C. and {Wallace}, Imaan E.~B. and {Williams}, Christina C. and {Willott}, Chris and {Whitler}, Lily and {Witstok}, Joris and {Woodrum}, Charity},
        title = "{JADES Initial Data Release for the Hubble Ultra Deep Field: Revealing the Faint Infrared Sky with Deep JWST NIRCam Imaging}",
      journal = {\apjs},
         year = 2023,
        month = nov,
       volume = {269},
       number = {1},
          eid = {16},
        pages = {16},
          doi = {10.3847/1538-4365/acf44d},
archivePrefix = {arXiv},
       eprint = {2306.02466},
 primaryClass = {astro-ph.GA},
       adsurl = {https://ui.adsabs.harvard.edu/abs/2023ApJS..269...16R}
}

@ARTICLE{Begelman_2006,
   author = {{Begelman}, M.~C. and {Volonteri}, M. and {Rees}, M.~J.},
    title = "{Formation of supermassive black holes by direct collapse in pre-galactic haloes}",
  journal = {\mnras},
   eprint = {astro-ph/0602363},
     year = 2006,
    month = jul,
   volume = 370,
    pages = {289-298},
      doi = {10.1111/j.1365-2966.2006.10467.x},
   adsurl = {http://adsabs.harvard.edu/abs/2006MNRAS.370..289B}
}

@ARTICLE{Modelevsky_2026,
       author = {{Modelevsky}, Elisha and {Nitzan}, Omri and {Sari}, Re'em and {Quataert}, Eliot},
        title = "{Emission line formation in scattering dominated media: implications for LRDs}",
      journal = {arXiv e-prints},
         year = 2026,
        month = aug,
          eid = {arXiv:2608.05283},
        pages = {arXiv:2608.05283},
archivePrefix = {arXiv},
       eprint = {2608.05283},
 primaryClass = {astro-ph.GA},
       adsurl = {https://ui.adsabs.harvard.edu/abs/2026arXiv260805283M}
}

@ARTICLE{Latif_2014_UV,
       author = {{Latif}, M.~A. and {Bovino}, S. and {Van Borm}, C. and {Grassi}, T. and {Schleicher}, D.~R.~G. and {Spaans}, M.},
        title = "{A UV flux constraint on the formation of direct collapse black holes}",
      journal = {\mnras},
         year = 2014,
        month = sep,
       volume = {443},
       number = {3},
        pages = {1979-1987},
          doi = {10.1093/mnras/stu1230},
archivePrefix = {arXiv},
       eprint = {1404.5773},
 primaryClass = {astro-ph.GA},
       adsurl = {https://ui.adsabs.harvard.edu/abs/2014MNRAS.443.1979L}
}

@ARTICLE{Omukai_2001,
       author = {{Omukai}, Kazuyuki},
        title = "{Primordial Star Formation under Far-Ultraviolet Radiation}",
      journal = {\apj},
         year = 2001,
        month = jan,
       volume = {546},
       number = {2},
        pages = {635-651},
          doi = {10.1086/318296},
archivePrefix = {arXiv},
       eprint = {astro-ph/0011446},
 primaryClass = {astro-ph},
       adsurl = {https://ui.adsabs.harvard.edu/abs/2001ApJ...546..635O}
}

@misc{Brammer_2023,
       author = {{Brammer}, Gabriel},
        title = "{grizli}",
         year = 2023,
        month = sep,
          eid = {10.5281/zenodo.8370018},
          doi = {10.5281/zenodo.8370018},
      version = {1.9.11},
      journal = {Zenodo},
    publisher = {Zenodo},
       adsurl = {https://ui.adsabs.harvard.edu/abs/2023zndo...8370018B}
}

@ARTICLE{Ji_2026a,
       author = {{Ji}, Zhiyuan and {Sun}, Yang and {Giavalisco}, Mauro and {Zhu}, Yongda and {Rieke}, George H. and {Williams}, Christina C. and {Maseda}, Michael V. and {Lyu}, Jianwei and {Rieke}, Marcia and {Tacchella}, Sandro},
        title = "{Compact Core, Extended Reach: A Bipolar Kiloparsec-scale Elongation in a Little Red Dot at z {\ensuremath{\approx}} 5.5}",
      journal = {\apjl},
         year = 2026,
        month = sep,
       volume = {1008},
       number = {2},
          eid = {L55},
        pages = {L55},
          doi = {10.3847/2041-8213/ae9fea},
archivePrefix = {arXiv},
       eprint = {2606.18342},
 primaryClass = {astro-ph.GA},
       adsurl = {https://ui.adsabs.harvard.edu/abs/2026ApJ..1008L..55J}
}

@ARTICLE{Calzetti_2000,
       author = {{Calzetti}, Daniela and {Armus}, Lee and {Bohlin}, Ralph C. and {Kinney}, Anne L. and {Koornneef}, Jan and {Storchi-Bergmann}, Thaisa},
        title = "{The Dust Content and Opacity of Actively Star-forming Galaxies}",
      journal = {\apj},
         year = 2000,
        month = apr,
       volume = {533},
       number = {2},
        pages = {682-695},
          doi = {10.1086/308692},
archivePrefix = {arXiv},
       eprint = {astro-ph/9911459},
 primaryClass = {astro-ph},
       adsurl = {https://ui.adsabs.harvard.edu/abs/2000ApJ...533..682C}
}

@ARTICLE{Ji_2026b,
       author = {{Ji}, Zhiyuan and {Sun}, Yang and {Giavalisco}, Mauro and {de Graaff}, Anna and {Williams}, Christina C. and {Zhu}, Yongda and {Rieke}, George H. and {Rieke}, Marcia},
        title = "{Through the Veil: Ly$α$ Illuminates the Host Galaxies of Little Red Dots}",
      journal = {arXiv e-prints},
         year = 2026,
        month = jun,
          eid = {arXiv:2606.09970},
        pages = {arXiv:2606.09970},
          doi = {10.48550/arXiv.2606.09970},
archivePrefix = {arXiv},
       eprint = {2606.09970},
 primaryClass = {astro-ph.GA},
       adsurl = {https://ui.adsabs.harvard.edu/abs/2026arXiv260609970J}
}

@ARTICLE{Deugenio_2025,
       author = {{D'Eugenio}, Francesco and {Nelson}, Erica and {Ji}, Xihan and {Baggen}, Josephine and {Greene}, Jenny and {Labb{\'e}}, Ivo and {Pezzulli}, Gabriele and {Brown}, Vanessa and {Maiolino}, Roberto and {Matthee}, Jorryt and {Terlevich}, Elena and {Terlevich}, Roberto and {Torralba}, Alberto and {Carniani}, Stefano},
        title = "{Irony at z=6.68: a bright AGN with forbidden Fe emission and multi-component Balmer absorption}",
      journal = {arXiv e-prints},
         year = 2025,
        month = sep,
          eid = {arXiv:2510.00101},
        pages = {arXiv:2510.00101},
          doi = {10.48550/arXiv.2510.00101},
archivePrefix = {arXiv},
       eprint = {2510.00101},
 primaryClass = {astro-ph.GA},
       adsurl = {https://ui.adsabs.harvard.edu/abs/2025arXiv251000101D}
}

@ARTICLE{Barger_2026,
       author = {{Barger}, A.~J. and {Cowie}, L.~L.},
        title = "{Double Dots: Compact Pairs Mark Little Red Dots and High-Redshift Broad-line AGNs}",
      journal = {arXiv e-prints},
         year = 2026,
        month = may,
          eid = {arXiv:2605.27903},
        pages = {arXiv:2605.27903},
archivePrefix = {arXiv},
       eprint = {2605.27903},
 primaryClass = {astro-ph.GA},
       adsurl = {https://ui.adsabs.harvard.edu/abs/2026arXiv260527903B}
}

@ARTICLE{Visbal_2014_pair,
       author = {{Visbal}, Eli and {Haiman}, Zolt{\'a}n and {Bryan}, Greg L.},
        title = "{Direct collapse black hole formation from synchronized pairs of atomic cooling haloes}",
      journal = {\mnras},
         year = 2014,
        month = nov,
       volume = {445},
       number = {1},
        pages = {1056-1063},
          doi = {10.1093/mnras/stu1794},
archivePrefix = {arXiv},
       eprint = {1406.7020},
 primaryClass = {astro-ph.GA},
       adsurl = {https://ui.adsabs.harvard.edu/abs/2014MNRAS.445.1056V}
}

@ARTICLE{Baggen_2026,
       author = {{Baggen}, Josephine F.~W. and {Scoggins}, Matthew T. and {van Dokkum}, Pieter and {Haiman}, Zolt{\'a}n and {Torralba}, Alberto and {Matthee}, Jorryt},
        title = "{Connecting the Dots: UV-bright Companions of Little Red Dots as Lyman─Werner Sources Enabling Direct-collapse Black Hole Formation}",
      journal = {\apjl},
         year = 2026,
        month = may,
       volume = {1002},
       number = {1},
          eid = {L4},
        pages = {L4},
          doi = {10.3847/2041-8213/ae58a5},
archivePrefix = {arXiv},
       eprint = {2602.02702},
 primaryClass = {astro-ph.GA},
       adsurl = {https://ui.adsabs.harvard.edu/abs/2026ApJ..1002L...4B}
}

@ARTICLE{Pacucci_2026_DCBH,
       author = {{Pacucci}, Fabio and {Ferrara}, Andrea and {Kocevski}, Dale D.},
        title = "{The Little Red Dots Are Direct Collapse Black Holes}",
      journal = {arXiv e-prints},
         year = 2026,
        month = jan,
          eid = {arXiv:2601.14368},
        pages = {arXiv:2601.14368},
          doi = {10.48550/arXiv.2601.14368},
archivePrefix = {arXiv},
       eprint = {2601.14368},
 primaryClass = {astro-ph.GA},
       adsurl = {https://ui.adsabs.harvard.edu/abs/2026arXiv260114368P}
}

@ARTICLE{Pacucci_Loeb_2025,
       author = {{Pacucci}, Fabio and {Loeb}, Abraham},
        title = "{Cosmic Outliers: Low-spin Halos Explain the Abundance, Compactness, and Redshift Evolution of the Little Red Dots}",
      journal = {\apjl},
         year = 2025,
        month = aug,
       volume = {989},
       number = {2},
          eid = {L19},
        pages = {L19},
          doi = {10.3847/2041-8213/ade871},
archivePrefix = {arXiv},
       eprint = {2506.03244},
 primaryClass = {astro-ph.GA},
       adsurl = {https://ui.adsabs.harvard.edu/abs/2025ApJ...989L..19P}
}

@ARTICLE{Rieke_2023,
       author = {{Rieke}, Marcia J. and {Kelly}, Douglas M. and {Misselt}, Karl and {Stansberry}, John and {Boyer}, Martha and {Beatty}, Thomas and {Egami}, Eiichi and {Florian}, Michael and {Greene}, Thomas P. and {Hainline}, Kevin and {Leisenring}, Jarron and {Roellig}, Thomas and {Schlawin}, Everett and {Sun}, Fengwu and {Tinnin}, Lee and {Williams}, Christina C. and {Willmer}, Christopher N.~A. and {Wilson}, Debra and {Clark}, Charles R. and {Rohrbach}, Scott and {Brooks}, Brian and {Canipe}, Alicia and {Correnti}, Matteo and {DiFelice}, Audrey and {Gennaro}, Mario and {Girard}, Julien H. and {Hartig}, George and {Hilbert}, Bryan and {Koekemoer}, Anton M. and {Nikolov}, Nikolay K. and {Pirzkal}, Norbert and {Rest}, Armin and {Robberto}, Massimo and {Sunnquist}, Ben and {Telfer}, Randal and {Wu}, Chi Rai and {Ferry}, Malcolm and {Lewis}, Dan and {Baum}, Stefi and {Beichman}, Charles and {Doyon}, Ren{\'e} and {Dressler}, Alan and {Eisenstein}, Daniel J. and {Ferrarese}, Laura and {Hodapp}, Klaus and {Horner}, Scott and {Jaffe}, Daniel T. and {Johnstone}, Doug and {Krist}, John and {Martin}, Peter and {McCarthy}, Donald W. and {Meyer}, Michael and {Rieke}, George H. and {Trauger}, John and {Young}, Erick T.},
        title = "{Performance of NIRCam on JWST in Flight}",
      journal = {\pasp},
         year = 2023,
        month = feb,
       volume = {135},
       number = {1044},
          eid = {028001},
        pages = {028001},
          doi = {10.1088/1538-3873/acac53},
archivePrefix = {arXiv},
       eprint = {2212.12069},
 primaryClass = {astro-ph.IM},
       adsurl = {https://ui.adsabs.harvard.edu/abs/2023PASP..135b8001R}
}

@ARTICLE{Maiolino_2024_Xray,
       author = {{Maiolino}, Roberto and {Risaliti}, Guido and {Signorini}, Matilde and {Trefoloni}, Bartolomeo and {Juod{\v{z}}balis}, Ignas and {Scholtz}, Jan and {{\"U}bler}, Hannah and {D'Eugenio}, Francesco and {Carniani}, Stefano and {Fabian}, Andy and {Ji}, Xihan and {Mazzolari}, Giovanni and {Bertola}, Elena and {Brusa}, Marcella and {Bunker}, Andrew J. and {Charlot}, Stephane and {Comastri}, Andrea and {Cresci}, Giovanni and {DeCoursey}, Christa Noel and {Egami}, Eiichi and {Fiore}, Fabrizio and {Gilli}, Roberto and {Perna}, Michele and {Tacchella}, Sandro and {Venturi}, Giacomo},
        title = "{JWST meets Chandra: a large population of Compton thick, feedback-free, and intrinsically X-ray weak AGN, with a sprinkle of SNe}",
      journal = {\mnras},
         year = 2025,
        month = apr,
       volume = {538},
       number = {3},
        pages = {1921-1943},
          doi = {10.1093/mnras/staf359},
archivePrefix = {arXiv},
       eprint = {2405.00504},
 primaryClass = {astro-ph.GA},
       adsurl = {https://ui.adsabs.harvard.edu/abs/2025MNRAS.538.1921M}
}

@ARTICLE{Kocevski_2024,
       author = {{Kocevski}, Dale D. and {Finkelstein}, Steven L. and {Barro}, Guillermo and {Taylor}, Anthony J. and {Calabr{\`o}}, Antonello and {Laloux}, Brivael and {Buchner}, Johannes and {Trump}, Jonathan R. and {Leung}, Gene C.~K. and {Yang}, Guang and {Dickinson}, Mark and {P{\'e}rez-Gonz{\'a}lez}, Pablo G. and {Pacucci}, Fabio and {Inayoshi}, Kohei and {Somerville}, Rachel S. and {McGrath}, Elizabeth J. and {Akins}, Hollis B. and {Bagley}, Micaela B. and {Bowler}, Rebecca A.~A. and {Bisigello}, Laura and {Carnall}, Adam and {Casey}, Caitlin M. and {Cheng}, Yingjie and {Cleri}, Nikko J. and {Costantin}, Luca and {Cullen}, Fergus and {Davis}, Kelcey and {Donnan}, Callum T. and {Dunlop}, James S. and {Ellis}, Richard S. and {Ferguson}, Henry C. and {Fujimoto}, Seiji and {Fontana}, Adriano and {Giavalisco}, Mauro and {Grazian}, Andrea and {Grogin}, Norman A. and {Hathi}, Nimish P. and {Hirschmann}, Michaela and {Huertas-Company}, Marc and {Holwerda}, Benne W. and {Illingworth}, Garth and {Juneau}, St{\'e}phanie and {Kartaltepe}, Jeyhan S. and {Koekemoer}, Anton M. and {Li}, Wenxiu and {Lucas}, Ray A. and {Magee}, Dan and {Mason}, Charlotte and {McLeod}, Derek J. and {McLure}, Ross J. and {Napolitano}, Lorenzo and {Papovich}, Casey and {Pirzkal}, Nor and {Rodighiero}, Giulia and {Santini}, Paola and {Wilkins}, Stephen M. and {Yung}, L.~Y. Aaron},
        title = "{The Rise of Faint, Red Active Galactic Nuclei at $z > 4$: A Sample of Little Red Dots in the JWST Extragalactic Legacy Fields}",
      journal = {\apj},
         year = 2025,
        month = jun,
       volume = {986},
       number = {2},
          eid = {126},
        pages = {126},
          doi = {10.3847/1538-4357/adbc7d},
archivePrefix = {arXiv},
       eprint = {2404.03576},
 primaryClass = {astro-ph.GA},
       adsurl = {https://ui.adsabs.harvard.edu/abs/2025ApJ...986..126K}
}

@ARTICLE{Pacucci_2023_JWST,
       author = {{Pacucci}, Fabio and {Nguyen}, Bao and {Carniani}, Stefano and {Maiolino}, Roberto and {Fan}, Xiaohui},
        title = "{JWST CEERS and JADES Active Galaxies at z = 4-7 Violate the Local $M_\bullet-M_\star$ Relation at $>3\sigma$: Implications for Low-mass Black Holes and Seeding Models}",
      journal = {\apjl},
         year = 2023,
        month = nov,
       volume = {957},
       number = {1},
          eid = {L3},
        pages = {L3},
          doi = {10.3847/2041-8213/ad0158},
archivePrefix = {arXiv},
       eprint = {2308.12331},
 primaryClass = {astro-ph.GA},
       adsurl = {https://ui.adsabs.harvard.edu/abs/2023ApJ...957L...3P}
}

@ARTICLE{Greene_2023,
       author = {{Greene}, Jenny E. and {Labbe}, Ivo and {Goulding}, Andy D. and {Furtak}, Lukas J. and {Chemerynska}, Iryna and {Kokorev}, Vasily and {Dayal}, Pratika and {Volonteri}, Marta and {Williams}, Christina C. and {Wang}, Bingjie and {Setton}, David J. and {Burgasser}, Adam J. and {Bezanson}, Rachel and {Atek}, Hakim and {Brammer}, Gabriel and {Cutler}, Sam E. and {Feldmann}, Robert and {Fujimoto}, Seiji and {Glazebrook}, Karl and {de Graaff}, Anna and {Khullar}, Gourav and {Leja}, Joel and {Marchesini}, Danilo and {Maseda}, Michael V. and {Matthee}, Jorryt and {Miller}, Tim B. and {Naidu}, Rohan P. and {Nanayakkara}, Themiya and {Oesch}, Pascal A. and {Pan}, Richard and {Papovich}, Casey and {Price}, Sedona H. and {van Dokkum}, Pieter and {Weaver}, John R. and {Whitaker}, Katherine E. and {Zitrin}, Adi},
        title = "{UNCOVER Spectroscopy Confirms the Surprising Ubiquity of Active Galactic Nuclei in Red Sources at z > 5}",
      journal = {\apj},
         year = 2024,
        month = mar,
       volume = {964},
       number = {1},
          eid = {39},
        pages = {39},
          doi = {10.3847/1538-4357/ad1e5f},
archivePrefix = {arXiv},
       eprint = {2309.05714},
 primaryClass = {astro-ph.GA},
       adsurl = {https://ui.adsabs.harvard.edu/abs/2024ApJ...964...39G}
}

@ARTICLE{Madau_2024,
       author = {{Madau}, Piero and {Haardt}, Francesco},
        title = "{X-Ray Weak Active Galactic Nuclei from Super-Eddington Accretion onto Infant Black Holes}",
      journal = {\apjl},
         year = 2024,
        month = dec,
       volume = {976},
       number = {2},
          eid = {L24},
        pages = {L24},
          doi = {10.3847/2041-8213/ad90e1},
archivePrefix = {arXiv},
       eprint = {2410.00417},
 primaryClass = {astro-ph.GA},
       adsurl = {https://ui.adsabs.harvard.edu/abs/2024ApJ...976L..24M}
}

@ARTICLE{Lambrides_2024,
       author = {{Lambrides}, Erini and {Larson}, Rebecca L. and {Garofali}, Kristen and {Ptak}, Andrew and {Chiaberge}, Marco and {Long}, Arianna S. and {Hutchison}, Taylor A. and {Norman}, Colin and {McKinney}, Jed and {Akins}, Hollis B. and {Berg}, Danielle A. and {Chisholm}, John and {Civano}, Francesca and {Cloonan}, Aidan P. and {Endsley}, Ryan and {Faisst}, Andreas L. and {Gilli}, Roberto and {Gillman}, Steven and {Hirschmann}, Michaela and {Kartaltepe}, Jeyhan S. and {Kocevski}, Dale D. and {Kokorev}, Vasily and {Pacucci}, Fabio and {Richardson}, Chris T. and {Stiavelli}, Massimo and {Whalen}, Kelly E.},
        title = "{The case for super-Eddington accretion in JWST broad-line active galactic nuclei during the first billion years}",
      journal = {Nature Astronomy},
         year = 2026,
        month = jun,
       volume = {10},
        pages = {868-879},
          doi = {10.1038/s41550-026-02813-w},
archivePrefix = {arXiv},
       eprint = {2409.13047},
 primaryClass = {astro-ph.HE},
       adsurl = {https://ui.adsabs.harvard.edu/abs/2026NatAs..10..868L}
}

@ARTICLE{Perez-Gonzalez_2026,
       author = {{P{\'e}rez-Gonz{\'a}lez}, Pablo G. and {Barro}, Guillermo and {Carniani}, Stefano and {D'Eugenio}, Francesco and {Rieke}, George H. and {Tripodi}, Roberta and {Bunker}, Andrew J. and {Ji}, Xihan and {Marques-Chaves}, Rui and {Schaerer}, Daniel and {Venturi}, Giacomo and {Ar{\'e}valo-Gonz{\'a}lez}, Flor and {Arribas}, Santiago and {Rinaldi}, Pierluigi and {Rodr{\'\i}guez Del Pino}, Bruno and {Witstok}, Joris and {Bhatawdekar}, Rachana and {Boogaard}, Leindert A. and {Charlot}, Stephane and {Chevallard}, Jacopo and {Costantin}, Luca and {Curti}, Mirko and {Curtis-Lake}, Emma and {Daddi}, Emanuele and {Davis}, Kelcey and {Dickinson}, Mark and {Donnan}, Callum T. and {Donnan}, Fergus R. and {Dunlop}, James S. and {Eisenstein}, Daniel J. and {Ferguson}, Henry C. and {Fern{\'a}ndez Aranda}, Rom{\'a}n and {Finkelstein}, Steven L. and {Fujimoto}, Seiji and {Gandolfi}, Giovanni and {Giavalisco}, Mauro and {Grogin}, Norman A. and {Hamed}, Mahmoud and {Hirschmann}, Michaela and {Kartaltepe}, Jeyhan S. and {Kocevski}, Dale D. and {Koekemoer}, Anton M. and {Leung}, Gene C.~K. and {Lofaro}, Cristina M. and {Lucas}, Ray A. and {McLeod}, Derek J. and {Melinder}, Jens and {{\"O}stlin}, Goran and {Papovich}, Casey and {Pentericci}, Laura and {P{\'e}rez-D{\'\i}az}, Borja and {Rieke}, Marcia and {Scholtz}, Jan and {Somerville}, Rachel S. and {Stanton}, Thomas M. and {Stevenson}, Struan D. and {Shivaei}, Irene and {Tacchella}, Sandro and {Trump}, Jonathan R. and {{\"U}bler}, Hannah and {Wang}, Xin and {Williams}, Christina C. and {Willmer}, Christopher N.~A. and {Yung}, L.~Y. Aaron and {Zhu}, Yongda},
        title = "{Little Red Dots: One Photometric Tag Concealing Diverse Spectroscopic Flavors of Massive Star Formation and Black Hole Activity}",
      journal = {arXiv e-prints},
         year = 2026,
        month = feb,
          eid = {arXiv:2602.20247},
        pages = {arXiv:2602.20247},
          doi = {10.48550/arXiv.2602.20247},
archivePrefix = {arXiv},
       eprint = {2602.20247},
 primaryClass = {astro-ph.GA},
       adsurl = {https://ui.adsabs.harvard.edu/abs/2026arXiv260220247P}
}

@ARTICLE{Akins_2025,
       author = {{Akins}, Hollis B. and {Casey}, Caitlin M. and {Lambrides}, Erini and {Allen}, Natalie and {Andika}, Irham T. and {Brinch}, Malte and {Champagne}, Jaclyn B. and {Cooper}, Olivia and {Ding}, Xuheng and {Drakos}, Nicole E. and {Faisst}, Andreas and {Finkelstein}, Steven L. and {Franco}, Maximilien and {Fujimoto}, Seiji and {Gentile}, Fabrizio and {Gillman}, Steven and {Gozaliasl}, Ghassem and {Harish}, Santosh and {Hayward}, Christopher C. and {Hirschmann}, Michaela and {Ilbert}, Olivier and {Kartaltepe}, Jeyhan S. and {Kocevski}, Dale D. and {Koekemoer}, Anton M. and {Kokorev}, Vasily and {Liu}, Daizhong and {Long}, Arianna S. and {McCracken}, Henry Joy and {McKinney}, Jed and {Onoue}, Masafusa and {Paquereau}, Louise and {Renzini}, Alvio and {Rhodes}, Jason and {Robertson}, Brant E. and {Shuntov}, Marko and {Silverman}, John D. and {Tanaka}, Takumi S. and {Toft}, Sune and {Trakhtenbrot}, Benny and {Valentino}, Francesco and {Zavala}, Jorge},
        title = "{COSMOS-Web: The Overabundance and Physical Nature of ``Little Red Dots''{\textemdash}Implications for Early Galaxy and SMBH Assembly}",
      journal = {\apj},
         year = 2025,
        month = sep,
       volume = {991},
       number = {1},
          eid = {37},
        pages = {37},
          doi = {10.3847/1538-4357/ade984},
archivePrefix = {arXiv},
       eprint = {2406.10341},
 primaryClass = {astro-ph.GA},
       adsurl = {https://ui.adsabs.harvard.edu/abs/2025ApJ...991...37A}
}

@ARTICLE{Maiolino_2023_new,
       author = {{Maiolino}, Roberto and {Scholtz}, Jan and {Curtis-Lake}, Emma and {Carniani}, Stefano and {Baker}, William and {de Graaff}, Anna and {Tacchella}, Sandro and {{\"U}bler}, Hannah and {D'Eugenio}, Francesco and {Witstok}, Joris and {Curti}, Mirko and {Arribas}, Santiago and {Bunker}, Andrew J. and {Charlot}, St{\'e}phane and {Chevallard}, Jacopo and {Eisenstein}, Daniel J. and {Egami}, Eiichi and {Ji}, Zhiyuan and {Jones}, Gareth C. and {Lyu}, Jianwei and {Rawle}, Tim and {Robertson}, Brant and {Rujopakarn}, Wiphu and {Perna}, Michele and {Sun}, Fengwu and {Venturi}, Giacomo and {Williams}, Christina C. and {Willott}, Chris},
        title = "{JADES: The diverse population of infant black holes at $4 < z < 11$: Merging, tiny, poor, but mighty}",
      journal = {\aap},
         year = 2024,
        month = nov,
       volume = {691},
          eid = {A145},
        pages = {A145},
          doi = {10.1051/0004-6361/202347640},
archivePrefix = {arXiv},
       eprint = {2308.01230},
 primaryClass = {astro-ph.GA},
       adsurl = {https://ui.adsabs.harvard.edu/abs/2024A&A...691A.145M}
}

@ARTICLE{Reines_2013,
       author = {{Reines}, Amy E. and {Greene}, Jenny E. and {Geha}, Marla},
        title = "{Dwarf Galaxies with Optical Signatures of Active Massive Black Holes}",
      journal = {\apj},
         year = 2013,
        month = oct,
       volume = {775},
       number = {2},
          eid = {116},
        pages = {116},
          doi = {10.1088/0004-637X/775/2/116},
archivePrefix = {arXiv},
       eprint = {1308.0328},
 primaryClass = {astro-ph.CO},
       adsurl = {https://ui.adsabs.harvard.edu/abs/2013ApJ...775..116R}
}

@ARTICLE{Hernquist_1990,
       author = {{Hernquist}, Lars},
        title = "{An Analytical Model for Spherical Galaxies and Bulges}",
      journal = {\apj},
         year = 1990,
        month = jun,
       volume = {356},
        pages = {359},
          doi = {10.1086/168845},
       adsurl = {https://ui.adsabs.harvard.edu/abs/1990ApJ...356..359H}
}

@ARTICLE{Inayoshi_review_2019,
       author = {{Inayoshi}, Kohei and {Visbal}, Eli and {Haiman}, Zolt{\'a}n},
        title = "{The Assembly of the First Massive Black Holes}",
      journal = {\araa},
         year = 2020,
        month = aug,
       volume = {58},
        pages = {27-97},
          doi = {10.1146/annurev-astro-120419-014455},
archivePrefix = {arXiv},
       eprint = {1911.05791},
 primaryClass = {astro-ph.GA},
       adsurl = {https://ui.adsabs.harvard.edu/abs/2020ARA&A..58...27I}
}

@ARTICLE{Astropy_2018,
       author = {{Astropy Collaboration} and {Price-Whelan}, A.~M. and {Sip{\H{o}}cz}, B.~M. and {G{\"u}nther}, H.~M. and {Lim}, P.~L. and {Crawford}, S.~M. and {Conseil}, S. and {Shupe}, D.~L. and {Craig}, M.~W. and {Dencheva}, N. and {Ginsburg}, A. and {VanderPlas}, J.~T. and {Bradley}, L.~D. and {P{\'e}rez-Su{\'a}rez}, D. and {de Val-Borro}, M. and {Aldcroft}, T.~L. and {Cruz}, K.~L. and {Robitaille}, T.~P. and {Tollerud}, E.~J. and {Ardelean}, C. and {Babej}, T. and {Bach}, Y.~P. and {Bachetti}, M. and {Bakanov}, A.~V. and {Bamford}, S.~P. and {Barentsen}, G. and {Barmby}, P. and {Baumbach}, A. and {Berry}, K.~L. and {Biscani}, F. and {Boquien}, M. and {Bostroem}, K.~A. and {Bouma}, L.~G. and {Brammer}, G.~B. and {Bray}, E.~M. and {Breytenbach}, H. and {Buddelmeijer}, H. and {Burke}, D.~J. and {Calderone}, G. and {Cano Rodr{\'\i}guez}, J.~L. and {Cara}, M. and {Cardoso}, J.~V.~M. and {Cheedella}, S. and {Copin}, Y. and {Corrales}, L. and {Crichton}, D. and {D'Avella}, D. and {Deil}, C. and {Depagne}, {\'E}. and {Dietrich}, J.~P. and {Donath}, A. and {Droettboom}, M. and {Earl}, N. and {Erben}, T. and {Fabbro}, S. and {Ferreira}, L.~A. and {Finethy}, T. and {Fox}, R.~T. and {Garrison}, L.~H. and {Gibbons}, S.~L.~J. and {Goldstein}, D.~A. and {Gommers}, R. and {Greco}, J.~P. and {Greenfield}, P. and {Groener}, A.~M. and {Grollier}, F. and {Hagen}, A. and {Hirst}, P. and {Homeier}, D. and {Horton}, A.~J. and {Hosseinzadeh}, G. and {Hu}, L. and {Hunkeler}, J.~S. and {Ivezi{\'c}}, {\v{Z}}. and {Jain}, A. and {Jenness}, T. and {Kanarek}, G. and {Kendrew}, S. and {Kern}, N.~S. and {Kerzendorf}, W.~E. and {Khvalko}, A. and {King}, J. and {Kirkby}, D. and {Kulkarni}, A.~M. and {Kumar}, A. and {Lee}, A. and {Lenz}, D. and {Littlefair}, S.~P. and {Ma}, Z. and {Macleod}, D.~M. and {Mastropietro}, M. and {McCully}, C. and {Montagnac}, S. and {Morris}, B.~M. and {Mueller}, M. and {Mumford}, S.~J. and {Muna}, D. and {Murphy}, N.~A. and {Nelson}, S. and {Nguyen}, G.~H. and {Ninan}, J.~P. and {N{\"o}the}, M. and {Ogaz}, S. and {Oh}, S. and {Parejko}, J.~K. and {Parley}, N. and {Pascual}, S. and {Patil}, R. and {Patil}, A.~A. and {Plunkett}, A.~L. and {Prochaska}, J.~X. and {Rastogi}, T. and {Reddy Janga}, V. and {Sabater}, J. and {Sakurikar}, P. and {Seifert}, M. and {Sherbert}, L.~E. and {Sherwood-Taylor}, H. and {Shih}, A.~Y. and {Sick}, J. and {Silbiger}, M.~T. and {Singanamalla}, S. and {Singer}, L.~P. and {Sladen}, P.~H. and {Sooley}, K.~A. and {Sornarajah}, S. and {Streicher}, O. and {Teuben}, P. and {Thomas}, S.~W. and {Tremblay}, G.~R. and {Turner}, J.~E.~H. and {Terr{\'o}n}, V. and {van Kerkwijk}, M.~H. and {de la Vega}, A. and {Watkins}, L.~L. and {Weaver}, B.~A. and {Whitmore}, J.~B. and {Woillez}, J. and {Zabalza}, V. and {Astropy Contributors}},
        title = "{The Astropy Project: Building an Open-science Project and Status of the v2.0 Core Package}",
      journal = {\aj},
         year = 2018,
        month = sep,
       volume = {156},
       number = {3},
          eid = {123},
        pages = {123},
          doi = {10.3847/1538-3881/aabc4f},
archivePrefix = {arXiv},
       eprint = {1801.02634},
 primaryClass = {astro-ph.IM},
       adsurl = {https://ui.adsabs.harvard.edu/abs/2018AJ....156..123A}
}

@ARTICLE{Matplotlib_2007,
       author = {{Hunter}, John D.},
        title = "{Matplotlib: A 2D Graphics Environment}",
      journal = {Computing in Science and Engineering},
         year = 2007,
        month = jan,
       volume = {9},
       number = {3},
        pages = {90-95},
          doi = {10.1109/MCSE.2007.55},
       adsurl = {https://ui.adsabs.harvard.edu/abs/2007CSE.....9...90H}
}

@ARTICLE{Scipy_2020,
       author = {{Virtanen}, Pauli and {Gommers}, Ralf and {Oliphant}, Travis E. and {Haberland}, Matt and {Reddy}, Tyler and {Cournapeau}, David and {Burovski}, Evgeni and {Peterson}, Pearu and {Weckesser}, Warren and {Bright}, Jonathan and {van der Walt}, St{\'e}fan J. and {Brett}, Matthew and {Wilson}, Joshua and {Millman}, K. Jarrod and {Mayorov}, Nikolay and {Nelson}, Andrew R.~J. and {Jones}, Eric and {Kern}, Robert and {Larson}, Eric and {Carey}, C.~J. and {Polat}, {\.I}lhan and {Feng}, Yu and {Moore}, Eric W. and {VanderPlas}, Jake and {Laxalde}, Denis and {Perktold}, Josef and {Cimrman}, Robert and {Henriksen}, Ian and {Quintero}, E.~A. and {Harris}, Charles R. and {Archibald}, Anne M. and {Ribeiro}, Ant{\^o}nio H. and {Pedregosa}, Fabian and {van Mulbregt}, Paul and {SciPy 1.  0 Contributors}},
        title = "{SciPy 1.0: fundamental algorithms for scientific computing in Python}",
      journal = {Nature Methods},
         year = 2020,
        month = feb,
       volume = {17},
        pages = {261-272},
          doi = {10.1038/s41592-019-0686-2},
archivePrefix = {arXiv},
       eprint = {1907.10121},
 primaryClass = {cs.MS},
       adsurl = {https://ui.adsabs.harvard.edu/abs/2020NaMet..17..261V}
}

@ARTICLE{Numpy_2020,
       author = {{Harris}, Charles R. and {Millman}, K. Jarrod and {van der Walt}, St{\'e}fan J. and {Gommers}, Ralf and {Virtanen}, Pauli and {Cournapeau}, David and {Wieser}, Eric and {Taylor}, Julian and {Berg}, Sebastian and {Smith}, Nathaniel J. and {Kern}, Robert and {Picus}, Matti and {Hoyer}, Stephan and {van Kerkwijk}, Marten H. and {Brett}, Matthew and {Haldane}, Allan and {del R{\'\i}o}, Jaime Fern{\'a}ndez and {Wiebe}, Mark and {Peterson}, Pearu and {G{\'e}rard-Marchant}, Pierre and {Sheppard}, Kevin and {Reddy}, Tyler and {Weckesser}, Warren and {Abbasi}, Hameer and {Gohlke}, Christoph and {Oliphant}, Travis E.},
        title = "{Array programming with NumPy}",
      journal = {\nat},
         year = 2020,
        month = sep,
       volume = {585},
       number = {7825},
        pages = {357-362},
          doi = {10.1038/s41586-020-2649-2},
archivePrefix = {arXiv},
       eprint = {2006.10256},
 primaryClass = {cs.MS},
       adsurl = {https://ui.adsabs.harvard.edu/abs/2020Natur.585..357H}
}

@ARTICLE{Bromm_Loeb_2003,
   author = {{Bromm}, V. and {Loeb}, A.},
    title = "{Formation of the First Supermassive Black Holes}",
  journal = {\apj},
   eprint = {astro-ph/0212400},
     year = 2003,
    month = oct,
   volume = 596,
    pages = {34-46},
      doi = {10.1086/377529},
   adsurl = {http://adsabs.harvard.edu/abs/2003ApJ...596...34B}
}
\bibliographystyle{aasjournalv7.1}



\end{document}